%% file: ms.tex
\documentclass{article}
\usepackage[utf8]{inputenc}
\RequirePackage{silence}
\usepackage{fullpage}
\usepackage{amsthm}
\usepackage{complexity}
\usepackage{mathtools,amsmath,amssymb}
\usepackage{graphicx}
\usepackage{color}
\usepackage{pgfplots}
\pgfplotsset{compat=1.7}
\usepackage{multirow}
\usepackage{tabularx}
\usepackage{bm}
\usepackage{enumitem}
\usepackage[colorinlistoftodos]{todonotes}
\usepackage[colorlinks=true, allcolors=blue]{hyperref}
\usepackage{pifont}

\newcolumntype{C}[1]{>{\centering\arraybackslash}p{#1}}

\renewcommand{\epsilon}{\varepsilon}
\newcommand{\eps}{\varepsilon}

\RequirePackage[colorlinks=true]{hyperref}
\hypersetup{
  linkcolor=[rgb]{0,0,0.5},
  citecolor=[rgb]{0, 0.5, 0},
  urlcolor=[rgb]{0.5, 0, 0}
}

\usepackage[breakable]{tcolorbox}
\newtcolorbox{reduction}[2][]
{
  breakable,
  colframe = gray!50,
  colback  = gray!10,
  coltitle = gray!10!black,
  before skip = 10pt,
  after skip = 10pt,
  title    = \textbf{#2},
  #1,
}
\newtcolorbox{graphview}[2][]
{
  breakable,
  colframe = black!30,
  colback  = black!0,
  coltitle = gray!10!black,
  before skip = 10pt,
  after skip = 10pt,
  title    = \textbf{#2},
  #1,
}

\usepackage{algorithmicx}
\usepackage{algorithm}
\usepackage[noend]{algpseudocode}
\newcounter{algsubstate}
\renewcommand{\thealgsubstate}{\alph{algsubstate}}

\algnewcommand\algorithmicinput{\textbf{Input:}}
\algnewcommand\Input{\item[\algorithmicinput]}

\algnewcommand\algorithmicoutput{\textbf{Output:}}
\algnewcommand\Output{\item[\algorithmicoutput]}

\algnewcommand\algorithmicgoal{\textbf{Goal:}}
\algnewcommand\Goal{\item[\algorithmicgoal]}

\input{thmmacros.tex}

\input{macros.tex}

\author{
Chi-Ning Chou\thanks{School of Engineering and Applied Sciences, Harvard University, Cambridge, Massachusetts, USA. Supported by NSF awards CCF 1565264 and CNS 1618026. Email: \texttt{chiningchou@g.harvard.edu}.}
\and Alexander Golovnev\thanks{School of Engineering and Applied Sciences, Harvard University, Cambridge, Massachusetts, USA. Supported by a Rabin  Postdoctoral Fellowship. Email: \texttt{alexgolovnev@gmail.com}.}
\and Santhoshini Velusamy\thanks{School of Engineering and Applied Sciences, Harvard University, Cambridge, Massachusetts, USA. Supported in part by a Simons 
Investigator Award and NSF Award CCF 1715187. Email: \texttt{svelusamy@g.harvard.edu}.}}

\title{Optimal Streaming Approximations for all Boolean Max-2CSPs and Max-$k$SAT}

\begin{document}
\date{}
\sloppy
\maketitle

\begin{abstract}
We prove tight upper and lower bounds on approximation ratios of all Boolean \mtwocsp{} problems in the streaming model. Specifically, for every type of \mtwocsp{} problem, we give an explicit constant~$\alpha$, s.t. for any~$\eps>0$ (i) there is an $(\alpha-\epsilon)$-streaming approximation using space $O(\log{n})$; and (ii) any $(\alpha+\epsilon)$-streaming approximation requires space $\Omega(\sqrt{n})$. This generalizes the celebrated work of [Kapralov, Khanna, Sudan SODA~2015; Kapralov, Krachun STOC~2019], who showed that the optimal approximation ratio for \mcut{} was $1/2$. 

Prior to this work, the problem of determining this ratio was open for all other \mtwocsp{}s. Our results are quite surprising for some specific \mtwocsp{}s. For the \mdcut{} problem, there was a gap between an upper bound of $1/2$ and a lower bound of $2/5$ [Guruswami, Velingker, Velusamy APPROX~2017]. We show that neither of these bounds is tight, and the optimal ratio for \mdcut{} is~$4/9$. We also establish that the tight approximation for \msat{} is $\sqrt{2}/2$, and for \mesat{} it is $3/4$. As a byproduct, our result gives a separation between space-efficient approximations for \msat{} and \mesat{}. This is in sharp contrast to the setting of polynomial-time algorithms with polynomial space, where the two problems are known to be equally hard to approximate. Finally, we prove that the tight streaming approximation for \mksat{} is $\sqrt{2}/2$ for every $k\geq2$.
\end{abstract}

\vfill
\thispagestyle{empty}
\newpage
\tableofcontents
\thispagestyle{empty}
\newpage
\pagenumbering{arabic}

\input{intro}
\input{prelims}
\input{ub}

\input{lb}
\input{theorem}

\input{maxksat}

\input{future}
\input{acknowledgement}

\bibliographystyle{alpha}
\bibliography{mybib}

\end{document}

%% file: thmmacros.tex
\usepackage{amsthm}
\usepackage{thmtools,thm-restate}

\numberwithin{equation}{section}
\declaretheoremstyle[bodyfont=\it,qed=\qedsymbol]{noproofstyle}

\declaretheorem[name=Observation,numbered=no]{observation*}

\declaretheorem[numberlike=equation]{theorem}

\declaretheorem[name=Theorem,numbered=no]{theorem*}

\declaretheorem[numberlike=equation]{lemma}
\declaretheorem[name=Lemma,numbered=no]{lemma*}

\declaretheorem[numberlike=equation]{corollary}
\declaretheorem[name=Corollary,numbered=no]{corollary*}

\declaretheorem[numberlike=equation]{proposition}
\declaretheorem[name=Proposition,numbered=no]{proposition*}

\declaretheorem[numberlike=equation]{claim}
\declaretheorem[name=Claim,numbered=no]{claim*}

\declaretheorem[name=Conjecture,numbered=no]{conjecture*}

\declaretheorem[name=Question,numbered=no]{question*}

\declaretheoremstyle[bodyfont=\it]{defstyle} 

\declaretheorem[numberlike=equation,style=defstyle]{definition}
\declaretheorem[unnumbered,name=Definition,style=defstyle]{definition*}

\declaretheorem[unnumbered,name=Example,style=defstyle]{example*}

\declaretheorem[unnumbered,name=Notation=defstyle]{notation*}

\declaretheorem[unnumbered,name=Construction,style=defstyle]{construction*}

\declaretheoremstyle[]{rmkstyle} 

\newtheorem*{remark}{Remark}



%% file: macros.tex
\newcommand{\bias}{\textsf{bias}}

\newcommand{\mcut}{\textsf{Max-CUT}}
\newcommand{\mdcut}{\textsf{Max-DICUT}}
\newcommand{\mcsp}{\textsf{Max-CSP}}

\newcommand{\mtwocsp}{\textsf{Max-2CSP}}
\newcommand{\mand}{\textsf{Max-2AND}}
\newcommand{\mt}{\textsf{Max-TR}}
\newcommand{\mor}{\textsf{Max-2OR}}
\newcommand{\meor}{\textsf{Max-2EOR}}
\newcommand{\meand}{\textsf{Max-2EAND}}
\newcommand{\mxor}{\textsf{Max-2XOR}}
\newcommand{\mexor}{\textsf{Max-2EXOR}}
\newcommand{\msat}{\textsf{Max-2SAT}}
\newcommand{\mSAT}{\textsf{Max-SAT}}
\newcommand{\mesat}{\textsf{Exact Max-2SAT}}
\newcommand{\mksat}{\textsf{Max-$k$SAT}}

\newcommand{\val}{\textsf{val}}

\newcommand{\cspt}{\textsf{TR}}
\newcommand{\cspand}{\textsf{AND}}
\newcommand{\cspeand}{\textsf{EAND}}
\newcommand{\cspxor}{\textsf{XOR}}

\newcommand{\cspor}{\textsf{OR}}

\newcommand{\yes}{\textbf{YES}}
\newcommand{\no}{\textbf{NO}}
\newcommand{\ono}{\overline{\textbf{NO}}}

\newcommand{\Exp}{\mathop{\mathbb{E}}}
\newcommand{\posvar}{\mathrm{pos}}
\newcommand{\negvar}{\mathrm{neg}}
\newcommand{\DBHP}{\textsf{DBHP}}

\newcommand{\ALG}{\textbf{ALG}}

\newcommand{\cA}{\mathcal{A}}
\newcommand{\cB}{\mathcal{B}}

\newcommand{\cD}{\mathcal{D}}
\newcommand{\cF}{\mathcal{F}}
\newcommand{\cG}{\mathcal{G}}
\renewcommand{\cP}{\mathcal{P}}

\newcommand{\N}{\mathbb{N}}
\newcommand{\cX}{\mathcal{X}}
\newcommand{\Real}{\mathbb{R}}

\DeclarePairedDelimiter\ceil{\lceil}{\rceil}

\newcommand{\GOOD}{\textsf{GOOD}}

%% file: intro.tex
\section{Introduction}
\label{sec:intro}
Maximum Boolean Constraint Satisfaction Problems, or $\mcsp$s, are a central class of optimization problems, including as special cases problems such as $\mcut$, \textsf{3SAT}, \textsf{Graph Coloring}, and \textsf{Vertex Cover}~\cite{creignou2008boolean}. Given a set of allowed predicates $\cF$, $\mcsp(\cF)$ is the optimization problem defined as follows. Every instance $\Psi$ of the problem consists of a set of Boolean variables $\cX$, and a set of constraints applied to them. Each constraint is a predicate from $\cF$ applied to the variables from $\mathcal{X}$ or their negations. The goal is to compute the maximum number of simultaneously satisfiable constraints. For example, \mksat{} is $\mcsp\left(\cF_{\cspor_{\leq k}}\right)$ where $\cF_{\cspor_{\leq k}}$ is the set of \cspor{} predicates on at most $k$ variables.

Schaefer’s famous dichotomy theorem~\cite{Sch78,thapper2016complexity} states that for any set of allowed predicates $\cF$, solving $\mcsp(\cF)$ \emph{exactly} is either in $\P$ or $\NP$-hard. However, the landscape of \emph{approximation} algorithms for $\mcsp$s is much more complex (see~\cite{MM17} and references therein).

The \mtwocsp{} problem---\mcsp{} where all constraints have length at most $2$---is the most studied case of \mcsp{}, and it generalizes many optimization problems on graphs. Starting with the seminal work of Goemans and Williamson~\cite{GW95}, a series of works~\cite{feige1995approximating,zwick2000analyzing,lewin2002improved} developed a $0.87401$-approximation algorithm for all \mtwocsp{}s, while under the $\P\neq\NP$ and Unique Games conjectures some \mtwocsp{}s do not admit $0.9001$- and $0.87435$-approximations, respectively~\cite{Has01,TSSW00,austrin2010towards}. 

In this paper, we follow the line of work~\cite{KK15,KKS15,KKSV17,GVV17,KK19,GT19} that studies the \textit{unconditional hardness} of approximating \mtwocsp{} through the lens of \textit{streaming algorithms}.
Over the last decade, there has been a lot of interest in designing algorithms for processing large streams of data using limited space (see~\cite{mcgregor2014graph,chakrabarti2015data} and references therein). The streaming model was formally defined in~\cite{alon1999space,henzinger1998computing}. 

A streaming algorithm for a \mtwocsp{} problem makes one pass through the list of constraints and uses space that is sub-linear (ideally, poly-logarithmic) in the input size.\footnote{In this work we focus on randomized streaming algorithms that make one pass over the input in a fixed (adversarial) order, and return the correct answer with probability $3/4$.} Since the algorithm is space bounded, it cannot even store an assignment to the input variables. Thus, a streaming algorithm is required to output an estimate of the maximum number of simultaneously satisfiable constraints. Specifically, for $\alpha\in[0,1]$, an $\alpha$-approximate streaming algorithm outputs a value $v$ for which the following two conditions hold with probability $3/4$: 
(i) there exists an assignment $\sigma$ satisfying at least $v$ constraints, and
(ii) $v\geq \alpha\cdot\val$, where $\val$ is the maximum number of simultaneously satisfiable constraints.

Prior to this work, the only \mtwocsp{} for which we knew the optimal streaming approximation factor was \mcut{}. \mcut{} asks us to find a bipartition of the $n$ vertices of an undirected graph that maximizes the number of edges crossing the partition—called the ``cut''. Note that \mcut{} corresponds to the $\mcsp(\cF)$ where $\cF$ contains the binary \cspxor{} predicate.\footnote{Although formally \mcut{} is a special case of  $\mxor$ where all constraints are of the form $x_i\oplus x_j=1$, it can be shown that these two problems are equivalent.} \cite{Zel11} shows that exact streaming algorithms for \mcut{} require quadratic space $\Omega(n^2)$. Since a random partition of a graph with $m$ edges has cut of expected size $m/2$, a trivial streaming algorithm $1/2$-approximates \mcut{} with $O(\log{m})$ space. It is also easy to see that for every $\epsilon>0$, it suffices to store $\widetilde{O}(n)$ random edges of the graph to compute a $(1-\epsilon)$-approximation of \mcut{}. 
A recent line of work~\cite{KKS15,KK15,KKSV17,KK19} shows that these two trivial bounds are optimal, \emph{i.e.}, any $(1/2+\epsilon)$-approximation algorithm requires linear space $\Omega(n)$. 

However, the case for directed graphs is not nearly so well understood.
In the \mdcut{} problem (another special case of \mtwocsp{}), given a directed graph, one needs to compute the maximum number of edges going from the first to the second part of the graph under any bipartition. While ~\cite{KK19,KKS15} rules out a $(1/2+\epsilon)$-approximation for \mdcut{} too, the trivial algorithm gives only a $1/4$-approximation here. \cite{GVV17} gives a $2/5$-approximation for \mdcut, still leaving a gap between the upper and lower bounds.

Even the hardness of \msat{} is not known in the streaming setting. Recall that in \msat{} the only allowed predicates are variables and pairwise \cspor{}s. A random assignment gives a $1/2$-approximation, and the classical $(\sqrt{5}-1)/2\approx0.61$-approximate algorithm of \cite{lieberherr1979complexity} can be implemented in $O(\log{n})$ space using $\ell_1$-sketching~\cite{Indyk,KNW10}. No non-trivial upper bounds are known for \msat{}.

\subsection{Our contribution}
In this work, we resolve a natural question about the approximation guarantees of streaming algorithms for~\emph{every} \mtwocsp{} problem.

Before presenting our results, we need a way to classify Boolean functions of two variables.
Let $f\colon\{0,1\}^2\to\{0,1\}$ be a function, then 
\begin{itemize}
\item $f$ is of $\cspt$-type, or trivial, if $f$ depends on at most one of its inputs (trivial functions are the two constant functions, and the four functions which depend on one of the inputs);
\item $f$ is of $\cspor$-type if the truth table of $f$ has exactly one 0 and three 1s;
\item $f$ is of $\cspxor$-type if $f$ depends on both inputs, and the truth table of $f$ has exactly two 0s and two 1s;
\item $f$ is of $\cspand$-type if the truth table of $f$ has exactly three 0s and one 1.
\end{itemize}

If a set of allowed predicates $\cF$ contains only constraints of a type $\Lambda\in\{\cspor,\cspxor,\cspand\}$, then the corresponding \mtwocsp{} problem is called \textsf{Max-2E$\Lambda$} ($2$-Exact-$\Lambda$, meaning that all constraints have length exactly $2$). If $\cF$ contains only $\Lambda$-type constraints and trivial constraints, then the corresponding \mtwocsp{} problem is called \textsf{Max-2$\Lambda$}.

We abuse notation by identifying a set of allowed predicates $\cF$ with the set of types of its predicates. Also, for a set $\cF=\{\Lambda\}$ containing one element, we write $\cF=\Lambda$. Therefore, a $\mcsp(\cF)$ problem is defined by  $\cF\subseteq\{\cspt,\cspor,\cspxor,\cspand\}$. Note that every $\mtwocsp$ problem corresponds to one such $\cF$. 

For every $\mcsp(\cF)$ problem, we give an explicit constant $\alpha_\cF$ such that $(\alpha_\cF-\eps)$-approximation can be computed in $O(\log{n})$ space, while $(\alpha_\cF+\eps)$-approximation requires space $\Omega(\sqrt{n})$, for every $\eps>0$.

\begin{restatable}{theorem}{thmmain}
\label{thm:main}
Let $\cF\subseteq\{\cspt, \cspor,\cspxor,\cspand\}$ be a set of allowed binary predicates. Let ${\alpha_\cF=\min_{\cG\subseteq \cF}\alpha_\cG}$, where $\alpha_\cG$ is given in Table~\ref{table:results}.

For every $\epsilon>0$, there exists an $(\alpha_\cF-\epsilon)$-approximate streaming algorithm for $\mcsp(\cF)$ that uses space $O(\epsilon^{-2}\log n)$. On the other hand, any $(\alpha_\cF+\epsilon)$-approximate streaming algorithm for $\mcsp(\cF)$ requires space~$\Omega(\sqrt{n})$.
\end{restatable}

\newcommand{\resultstable}{
\def\arraystretch{1.5}
\begin{tabular}{|C{.10\textwidth}|C{.08\textwidth}|C{.08\textwidth}|C{.1\textwidth}|}
\hline
\multirow{2}{*}{Type $\cG$} & Tight bound & \multicolumn{2}{c|}{Previous bound} \\ \cline{2-4}
& $\alpha_\cG$   &  $\alpha^{\text{pr}}_\cG$ & Reference  \\ \hline
$\cspt$   & 1                  & 1 & Folklore \\ \hline
$\cspor$  & $\frac{3}{4}$         & $[\frac{3}{4},1]$ &  Folklore \\ \hline
$\{\cspt,\cspor\}$   & $\frac{\sqrt{2}}{2}$  & $[\frac{\sqrt{5}-1}{2},1]$ &  \cite{lieberherr1979complexity} \\ \hline
$\cspxor$ & $\frac{1}{2}$        & $\frac{1}{2}$ &   \cite{KK19} \\ \hline
$\cspand$  & $\frac{4}{9}$         & $[\frac{2}{5},\frac{1}{2}]$ & \cite{GVV17} \\ \hline
\end{tabular}
}

\begin{table}[ht]
\centering
\resultstable
\caption{Summary of known and new approximation factors $\alpha_\cG$ for  $\mcsp(\cG)$. We have suppressed~$(1\pm\epsilon)$ multiplicative factors.}
\label{table:results}
\end{table}

\vspace{-0.5cm}
\paragraph{Discussion.}
Interestingly, Theorem~\ref{thm:main} identifies five \mtwocsp{} problems which completely characterize the hardness of any \mtwocsp{} problem. Namely, we show that $\mcsp(\cF)$ is precisely as hard to approximate as the hardest of the problems from Table~\ref{table:results} expressible by predicates from $\cF$.

In particular, Theorem~\ref{thm:main} closes the gap between $2/5$~\cite{GVV17} and $1/2$~\cite{KKS15} for the streaming approximation ratio of \mdcut{}. We prove that neither of these bounds is tight, and that the correct bound is~$4/9$.\footnote{While Theorem~\ref{thm:main} states the bound for the \mand{} problem only, it is easy to see that the proof in Section~\ref{sec:reduction and} gives the same bound even for \mdcut{}.} Similarly, it shows that the $(\sqrt{5}-1)/2$-approximate algorithm of~\cite{lieberherr1979complexity} for \msat{} can be improved further, and that the optimal approximation ratio is $\sqrt{2}/2$.

Many streaming problems have space-accuracy tradeoffs allowing for better approximations with more space (\emph{e.g.},~\cite{chakrabarti2015data,assadi2016tight}). Curiously, Theorem~\ref{thm:main} shows that every $\mtwocsp(\cF)$ problem exhibits sharp threshold behavior: it needs only logarithmic space to be approximated up to some constant $\alpha_\cF$, and it requires polynomial space for every larger approximation factor.

In the classical setting, approximation algorithms for \mcsp{}s use space-inefficient techniques including semidefinite and linear programming, and network flow computations~\cite{Y94,goemans1994new, GW95,Has08, raghavendra2008optimal, raghavendra2010graph, MM17}. On the other hand, the best streaming algorithms for \mcsp{}s (except for the work~\cite{GVV17}) used only random assignments to the variables of the instance, including \mcut{}, \msat{}, and Unique Games problems. We design streaming algorithms for the \mand{} and \mor{} problems (\emph{i.e.}, $\cF=\{\cspt{},\cspand\}$ and $\cF=\{\cspt{},\cspor\}$) which significantly improve on the approximation ratios guaranteed by a~random assignment to the variables.  

Additionally, Theorem~\ref{thm:main} reveals a curious difference between streaming approximation of the cases $\cG=\cspor$ and $\cG=\{\cspt,\cspor\}$ (\emph{i.e.}, \mesat{} and \msat{}). The former problem can be $3/4$-approximated, while the latter does not admit better than $\sqrt{2}/2$-approximations. This shows that adding trivial constraints to \mesat{} actually makes the problem harder to approximate. This is in sharp contrast to the classical setting of polynomial-time algorithms with polynomial space, where approximation-preserving reductions between the two problems are known~\cite{Y94}. While $3/4$-approximation for \mesat{} is trivial, many $3/4$-approximation algorithms for \msat{} use non-efficient (though polynomial) linear programming routines. This led Williamson to pose a question in 1998 whether there exists an algorithm for \msat{} which does not use linear programming and at least matches the trivial $3/4$-approximation guarantee for \mesat~\cite{williamson1999lecture}. The affirmative answer to this question was given by Poloczek and Schnitger in 2011~\cite{poloczek2011randomized,poloczek2011bounds,van2011simpler,poloczek2017greedy}. Theorem~\ref{thm:main} complements this result by showing that there is no $\sqrt{2}/2<3/4$-approximation for \msat{} in the streaming setting, thus, separating space-efficient approximations for \msat{} and \mesat{}.

Our final contribution is a tight bound on the approximation ratio of streaming algorithms for all {\mksat{}} problems. We generalize the $\sqrt{2}/2$-approximation algorithm for \msat{} from Theorem~\ref{thm:main} to an algorithm for \mSAT{}, and a matching hardness result trivially follows from the hardness of \msat{}.
\begin{restatable}{theorem}{thmmksat}
\label{thm:mksat}
For every $\epsilon>0$, there exists an $(\sqrt{2}/2-\epsilon)$-approximate streaming algorithm for \mSAT{} that uses space $O(\epsilon^{-2}\log n)$. On the other hand, for any $k\geq2,\epsilon>0$ any $(\sqrt{2}/2+\epsilon)$-approximate streaming algorithm for \mksat{} requires space~$\Omega(\sqrt{n})$.
\end{restatable}

\subsection{Related Work}
\paragraph{Classical setting.}
For every $\mtwocsp(\cF)$ problem, a random assignment satisfies in expectation a constant fraction $\alpha^{\text{tr}}_\cF$ of the constraints (this algorithm can be easily derandomized via the method of conditional expectations). In particular, this algorithm gives $1/2$- and $1/4$-approximations for \mcut{} and \mtwocsp{}. On one hand, H\aa stad~\cite{Has01} used the PCP theorem to show that some \mcsp{} problems, \textit{e.g.}, \textsf{MAX-E3SAT}, do not admit better than $\alpha^{\text{tr}}_\cF$-approximations unless $\P=\NP$. On the other hand, Goemans and Williamson~\cite{GW95} used semidefinite programming (SDP) to significantly improve the bounds for \mcut{} and \mtwocsp{} to $0.87856$ and $0.79607$. H\aa stad~\cite{Has08} proved that there is an SDP-based approximation algorithm with a better than $\alpha^{\text{tr}}_\cF$ approximation guarantee for every \mtwocsp{}. Many of the SDP-based approximation algorithms are optimal under the Unique Games Conjecture~\cite{KV05,khot2007optimal}. We refer the reader to~\cite{MM17} for an up-to-date overview of the literature.

\paragraph{Streaming setting.}
While there is a trivial $1/2$-approximation for \mcut{} using space $O(\log n)$, Kapralov \emph{et al.}~\cite{KKS15} showed that for any constant $\epsilon>0$, a $(1/2+\epsilon)$-approximation requires space~$\tilde{\Omega}(\sqrt{n})$. Independently, Kogan and Krauthgamer~\cite{KK15} showed that (i) $(1-\epsilon)$-approximation requires space $\Omega(n^{1-\epsilon})$ and (ii) $4/5$-approximation requires $\Omega(n^{\tau})$ space for some constant $\tau>0$. In a subsequent work,~\cite{KKSV17} showed that $(1-\epsilon)$-approximation requires $\Omega(n)$ space. This line of work culminated in a recent result by Kapralov and Krachun~\cite{KK19} showing that any $(1/2+\epsilon)$-approximation for \mcut{} requires $\Omega(n)$ space.

Recently Guruswami \emph{et al.}~\cite{GVV17} gave a $(2/5-\epsilon)$-approximate algorithm for $\mdcut$ for any constant $\epsilon>0$, significantly improving on the trivial $1/4$-approximation. For $k$-\textsf{SAT}, Kogan and Krauthgamer~\cite{KK15} showed that there is a $(1-\epsilon)$-approximation using $\tilde{O}(\eps^{-2}kn)$ space. The hardness side has been widely open prior to this work and, to the best of our knowledge, the only other hardness result is by Guruswami and Tao~\cite{GT19} showing that $(1/p+\epsilon)$-approximation for Unique Games with alphabet size $p$ requires $\tilde{\Omega}(\sqrt{n})$ space for any constant $\epsilon>0$.

\subsection{Techniques}
\paragraph{Streaming algorithms.}
The first step of our proof of Theorem~\ref{thm:main} is two new algorithms for \mor{} and \mand{} that improve on the naive approximations for these problems. For these algorithms, we generalize the notion of bias~\cite{GVV17} to all \mtwocsp{} problems, and prove a series of bounds on the value of \mtwocsp{} w.r.t. the bias (and the numbers of trivial and non-trivial constraints in the instance). This results in log-space streaming algorithms that sketch the bias (and some additional information about the instance), and compute good estimates of the value of the instance.

It is not hard to see that \mand{} is the ``hardest'' \mtwocsp{} problem, \emph{i.e.}, an $\alpha$-approximation for \mand{} implies $\alpha$-approximations for all \mtwocsp{}s (see Section~\ref{sec:theorem}). Therefore, the hardness result of~\cite{KK19} for \mcut{} holds for \mand{} as well, ruling out the possibility of $(1/2+\eps)$-approximations. On the other hand, a random assignment for \mand{} formulas only guarantees a $1/4$-approximation. A~recent work~\cite{GVV17} improves the approximation ratio to $(2/5-\epsilon)$ as follows.

\begin{figure}[ht]
	\centering
	\begin{tikzpicture}[scale=0.95]
        \begin{axis}[
            xmin=0,xmax=1,ymin=0,ymax=1,xlabel=$\bias(\Psi)$, ylabel=Range of $\val(\Psi)$,
            xtick={0,0.2,0.4,0.6,0.8,1},
            xticklabels={$0$,$0.2m$,$0.4m$,$0.6m$,$0.8m$,$m$},
            ytick={0,0.2,0.4,0.6,0.8,1},
            yticklabels={$0$,$0.2m$,$0.4m$,$0.6m$,$0.8m$,$m$},
            ]
            \addplot[blue, ultra thick,smooth] {1/2+x/2};
            \addplot[red, ultra thick,samples=100] {max(x,0.25)};
            \addplot[green,ultra thick,smooth,domain=0:0.33] {0.25+x*x/(4*(1-2*x))};
        \end{axis}
  \end{tikzpicture}
  \caption{Upper and lower bounds on the maximum number $\val$ of simultaneously satisfiable constraints of $\mand$ as a function of $\bias$. The blue line is the upper bound $\frac{m+\bias}{2}$, and the red line is the lower bound $\max\left(\frac{m}{4},\bias\right)$ from~\cite{GVV17} (see Lemma~\ref{lem:GVV17}). The green line is the new lower bound $\frac{m}{4}+\frac{\bias^2}{4(m-2\bias)}$ from Lemma~\ref{lem:and bias lower bound} in the interval $\bias\in[0,m/3]$.}
  \label{fig:alg and}
\end{figure}
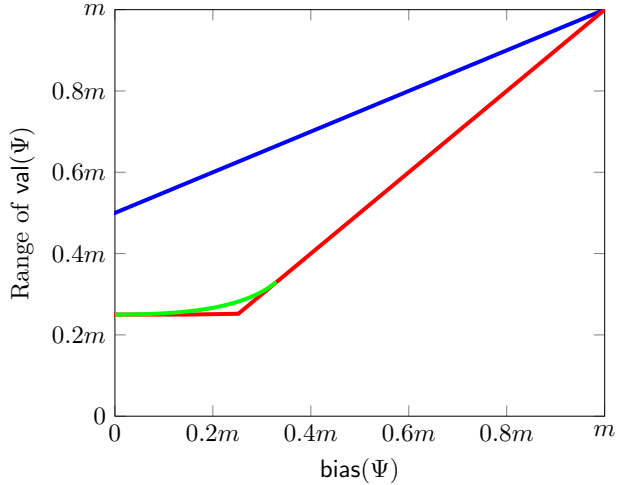

Let $\Psi$ be a $\meand{}$ instance with $m$ constraints, and $\val$ be the maximum number of simultaneously satisfiable constraints in $\Psi$.
\cite{GVV17} defines the bias of a variable $x$ as the absolute difference between the number of positive and negative occurrences of $x$, and the bias of the instance as the sum of biases of its variables. It is easy to see that for every instance, $\val\leq (m+\bias)/2$. \cite{GVV17} proves that the assignment of the input variables according to their biases satisfies at least $\bias$ constraints (see Lemma~\ref{lem:GVV17}). Then they conclude that $\max(\bias,m/4)$ is a $2/5$-approximation of $\val$:
\[
\frac{\max(\bias,m/4)}{\val} \geq \frac{\bias/5+(m/4)(4/5)}{(m+\bias)/2}=2/5
\, .
\]
The upper and lower bounds of~\cite{GVV17} are shown in red and blue in Figure~\ref{fig:alg and}, and the gap between the bounds indeed achieves $2/5$ when $\bias=m/4$. While both lower bounds $\val\geq\max(\bias,m/4)$ are tight as functions of $\bias$ and $m$, we show that in the important regime of low $\bias\in[0,m/3]$, these bounds can be improved to
\begin{align}
\label{eq:overview}
\val \geq \frac{m}{4}+\frac{\bias^2}{4(m-2\bias)} \, .
\end{align}
Unlike the lower bound of $\val\geq \bias$ from~\cite{GVV17}, our lower bound cannot be achieved by a greedy assignment to the input variables. Instead, we design a distribution of assignments, whose expected value is at least~\eqref{eq:overview}.
This improved lower bound on $\val$ (shown in green in Figure~\ref{fig:alg and}) leads to a $4/9$-approximation by a sketch for the expression~\eqref{eq:overview}. Namely, we give a $O(\log{n})$-space streaming algorithm that approximates the green and red bounds in Figure~\ref{fig:alg and}, and returns their maximum.

Perhaps surprisingly, the optimal approximation ratio for \mor{} significantly differs from both the $3/4$-approximation for \meor{}, and the trivial $1/2$-approximation. The classical algorithm of~\cite{lieberherr1979complexity} can be implemented in the streaming setting, but it only provides a $(\sqrt{5}-1)/2\approx0.61$-approximation. We prove that the tight bound for \mor{} is even larger---$\sqrt{2}/2$. Proofs of these upper and lower bounds are perhaps the most technical parts of this work. It can be shown that various naive random assignments to the variables used in $1/2$- and $(\sqrt{5}-1)/2$-approximations cannot lead to better bounds. Instead we construct a family of distributions of assignments which depend on individual biases of the variables. We use these distributions to prove the existence of assignments of some high value $v$, and finally we show a way to approximate $v$ in logarithmic space. We remark that it is not always possible to satisfy $m\sqrt{2}/2$ constraints, thus, we also prove non-trivial upper bounds on $\val$ for the case when our estimate $v$ is low $v<m\sqrt{2}/2$. (See Lemma~\ref{lem:ub2} and Lemma~\ref{lem:or bias lower bound} for formal statements of these results.)

\paragraph{Hardness results.}
We develop a framework for proving hardness results for various \mtwocsp{} problems, and use it to establish tight bounds for every $\mtwocsp$. This framework is based on the communication complexity lower bound of~\cite{KKS15} for the Distributional Boolean Hidden Partition problem ($\DBHP$) (which, in turn, extends the results of~\cite{gavinsky2007exponential,verbin2011streaming} for Boolean Hidden Matching and Boolean Hidden Hypermatching). In $\DBHP$, Alice holds a random bipartition of $[n]$, and Bob has a (random) graph $G$ on $n$ vertices with some edges marked. Their goal is to use minimal communication to distinguish between the following two cases: in the $\yes$ case, the set of Bob's marked edges is exactly the edges of $G$ that cross Alice's bipartition; while in the $\no$ case, a random subset of the edges is marked. \cite{KKS15} proved a lower bound of $\Omega(\sqrt{n})$ on the randomized one-way communication complexity of $\DBHP$.

For a set of allowed predicates $\cG$, we construct a reduction from $\DBHP$ to $\mcsp(\cG)$, which naturally induces distributions $\cD^Y$ and $\cD^N$ of $\mcsp(\cG)$ instances. Then by a careful analysis we show that the gap between the optimal solutions of instances from $\cD^Y$ and $\cD^N$ achieves $\alpha_{\cG}+\eps$ with high probability. This, amplified by a series of repetitions, lets us conclude that a space-efficient $(\alpha_{\cG}+\eps)$-approximate algorithm would contradict the lower bound on the communication complexity of $\DBHP$.

In our framework, we give separate reductions from $\DBHP$ to \meand{} and \mor{} with approximation ratios $4/9+\eps$ and $\sqrt{2}/2+\eps$, respectively. For the \meor{} problem, we give an efficient streaming reduction from \mcut{} to \meor{} which asserts that an $\alpha$-approximation for \meor{} implies an $\alpha/(3-2\alpha)$-approximation for \mcut{}. This, equipped with the lower bound from~\cite{KK19}, proves a \emph{linear} lower bound $\Omega(n)$ on the space complexity of $(3/4+\eps)$-approximations of \meor{}. 

\paragraph{Putting it all together.}
Finally, we show that our algorithms for the five problems from Table~\ref{table:results} can be combined together to handle every $\mtwocsp$ problem. Similarly, we prove that the established lower bounds for these five problems cover all possible \mtwocsp{}s. This implies that every $\mtwocsp$ problem $\mcsp(\cF)$ is precisely as hard to approximate as the hardest problem from \mt{}, \meor{}, \mor{}, \mexor{}, \meand{} expressible in $\cF$, and finishes the proof of Theorem~\ref{thm:main}.

\subsection{Structure}
In Section~\ref{sec:prelims}, we review some necessary background knowledge. In Section~\ref{sec:ub}, we provide streaming algorithms with optimal approximation ratios for all \mtwocsp{} problems. Sections~\ref{sec:lb} and~\ref{sec:gap instances} are devoted to proving tight bounds on the approximation ratios of streaming algorithms from Section~\ref{sec:ub}. In particular, Section~\ref{sec:lb} contains the general framework for our lower bounds, and the reductions from Distributional Boolean Hidden Partition to \mtwocsp{} problems. Section~\ref{sec:gap instances} provides a tight analysis of the approximation ratios resulting from these reductions. In Section~\ref{sec:theorem}, we combine the results of the previous sections to prove Theorem~\ref{thm:main}. Finally, Section~\ref{sec:mksat} gives an optimal streaming approximation algorithm for \mSAT{} and finishes the proof of Theorem~\ref{thm:mksat}.

%% file: prelims.tex
\section{Preliminaries}\label{sec:prelims}

Let $\mathbb{N}=\{1,2,\dots,\}$ be the set of natural numbers, and $[n]=\{1,2,\dots,n\}$ for any $n\in\mathbb{N}$. We use $\sqcup$ for the disjoint union of two sets. For an $0<\epsilon<1$, $B\in(1\pm\epsilon)$ is shorthand for $1-\epsilon\leq B\leq 1+\epsilon$. For ease of exposition we will abuse notation and associate a vector $X\in\{0,1\}^n$ with the set $X\subseteq [n], X=\{i\colon X_i=1\}$.

As we explained in Section~\ref{sec:intro}, we will primarily consider $\mcsp(\cG)$ where $\cG\in\{\cspt, \cspor, \{\cspt, \cspor\}, \cspxor, \cspand\}$. In order to get familiar with these problems, we provide several examples in Table~\ref{table:examples}.

\begin{table}[H]
\def\arraystretch{2}
\centering
\begin{tabular}{|C{.16\textwidth}|C{.16\textwidth}|C{.16\textwidth}|C{.16\textwidth}|C{.16\textwidth}|}
\hline
type $\cG$  & \cspor{} & $\{\cspt,\cspor\}$ & \cspxor{} & \cspand{} \\
\hline
problem name &  \meor{} & \mor{} & \mexor{} & \meand{}  \\ 
\hline
special case & \mesat{} & \msat{} & \mcut{} & \mdcut{}   \\ 
\hline
\end{tabular}
\caption{For each case $\cG\in\{\cspor, \{\cspt, \cspor\}, \cspxor, \cspand\}$, we give the name of the $\mcsp(\cG)$ problem, as~well as one well-studied special case/alternative name of the problem.}
\label{table:examples}
\end{table}

For an instance $\Psi$ of a \mtwocsp{} problem, we denote the number of clauses (constraints) in $\Psi$ by $m=|\Psi|$. We denote the set of Boolean variables of $\Psi$ by $\cX=\{x_1,\ldots,x_n\}$. A literal $\ell$ is called positive if $\ell=x_i$, and negative if $\ell=\neg x_i$ for some variable $x_i$. A $1$-clause is a clause (constraint) which depends only on one variable. We use $\posvar_i^{(1)}(\Psi)$ and $\posvar_i^{(2)}(\Psi)$ for the number of 1- and 2-clauses where the variable $x_i$ appears positively. Similarly, $\negvar_i^{(1)}(\Psi)$ and $\negvar_i^{(2)}(\Psi)$ denote the number of 1- and 2-clauses containing $\neg x_i$.

For an assignment $\sigma:\cX\rightarrow \{0,1\}$ of the variables of $\Psi$, we denote the number of clauses of $\Psi$ satisfied by $\sigma$ as $\val_\Psi(\sigma)$. We denote the maximum number of simultaneously satisfiable clauses in $\Psi$ as $\val_\Psi$:
\[
 \val_\Psi=\max_\sigma{\val_\Psi(\sigma)} \, .
\]

For $\alpha\in[0,1]$ and a set of allowed predicates $\cF$, an algorithm $\cA$ is an $\alpha$-approximation to the $\mcsp(\cF)$ problem if on any input $\Psi$, $\cA$ outputs $v$, such that with probability $3/4$, it holds that
\(
\val_\Psi \geq v \geq\alpha\cdot\val_\Psi \, .
\)
For example, when $\alpha=1$, the algorithm solves $\mcsp(\cF)$ exactly (with probability $3/4)$.
 
We will use the following definition of the bias of $\Psi$, which generalizes the definition from~\cite{GVV17} to all \mtwocsp{}s with clauses of length 1 or 2.\footnote{For uniformity reasons, our definition of bias differs from the definition in~\cite{GVV17} by a multiplicative factor of 2.}

\begin{definition}[Bias] \label{def:biasinstance}
The \emph{bias} of a variable $x_i$ of an instance $\Psi$ is defined as 
\[
\bias_i(\Psi) = \frac{1}{2}\cdot |2\posvar_i^{(1)}(\Psi)+\posvar_i^{(2)}(\Psi) -2\negvar_i^{(1)}(\Psi) -\negvar_i^{(2)}(\Psi)|
\, .
\]
The {bias vector} of $\Psi$ is a vector $\bm{b}\in\Real^n$, where $\bm{b}_i=\bias_i(\Psi)$.
Finally, the \emph{bias} of the formula $\Psi$ is defined as the sum of biases of its variables:
\[
 \bias(\Psi) = \sum_{i=1}^n \bias_i(\Psi) = \frac{1}{2}\sum_{i=1}^n |2\posvar_i^{(1)}(\Psi)+\posvar_i^{(2)}(\Psi) -2\negvar_i^{(1)}(\Psi) -\negvar_i^{(2)}(\Psi)| \, .
\]
\end{definition}

Note that for every formula $\Psi$ with $|\Psi|=m$ clauses, $0\leq \bias(\Psi) \leq m$.

In order to approximate the bias of a formula $\Psi$, we will use a streaming algorithm for approximating the~$\ell_1$ norm of the bias vector of $\Psi$.
\begin{theorem}[\cite{Indyk,KNW10}]
\label{thm:l1sketch}
Given a stream $S$ of $\poly(n)$ updates $(i, v) \in [n] \times \{1,-1\}$, let $x_i=\sum_{(i,v)\in S} v$ for $i\in[n]$. There exists a $1$-pass streaming algorithm, which uses $O(\log{n}/\epsilon^2)$ bits of memory and outputs a $(1\pm\epsilon)$-approximation to the value $\ell_1(x)=\sum_i |x_i|$ with probability $3/4$.
\end{theorem}

We will need the following concentration inequality~\cite{KK19}.

\begin{lemma}[{\cite[Lemma~2.5]{KK19}}]\label{lem:chernoff}
Let $X=\sum_{i\in[N]}X_i$ where $X_i$ are Bernoulli random variables such that for any $k\in[N]$, $\Exp[X_k|X_1,\dots,X_{k-1}]\leq p$ for some $p\in(0,1)$. Let $\mu=Np$. For any $\Delta>0$,
\[
\Pr\left[X\geq\mu+\Delta\right]\leq\exp\left(-\frac{\Delta^2}{2\mu+2\Delta}\right) \, .
\]
\end{lemma}

Finally, we will use the lower bound on the space complexity of streaming algorithms for approximate \mcut{} from~\cite{KK19}.
\begin{theorem}
\label{thm:kk19}
For any constant $\epsilon>0$, any streaming algorithm that $(1/2+\epsilon)$-approximates \mcut{} with success probability at least $3/4$ requires $\Omega(n)$ space.
\end{theorem}

\subsection{Total variation distance}
\begin{definition}[Total variation distance of discrete random variables]
Let $\Omega$ be a finite probability space and $X,Y$ be random variables with support $\Omega$. The total variation distance between $X$ and $Y$ is defined as follows.
\[
\|X-Y\|_{tvd} :=\frac{1}{2} \sum_{\omega\in\Omega}\left|\Pr[X=\omega]-\Pr[Y=\omega]\right| \, .
\]
\end{definition}

We will use the two following properties of the total variation distance.
\begin{proposition}\label{lem:tvd properties}
Let $\Omega$ be a finite probability space and $X,Y$ be random variables with support $\Omega$.
\begin{enumerate}
\item (Triangle inequality) Let $W$ be an arbitrary random variable, then we have $\|X-Y\|_{tvd}\geq\|X-W\|_{tvd}-\|Y-W\|_{tvd}$.
\item (Data processing inequality) Let $W$ be a random variable that is independent of both $X$ and $Y$, and $f$ be a function, then we have $\|f(X,W)-f(Y,W)\|_{tvd}\leq\|X-Y\|_{tvd}$.
\end{enumerate}
\end{proposition}
The triangle inequality for the total variation distance is a standard fact; and the proof of the data processing inequality can be found in~{\cite[Claim~6.5]{KKS15}}.

%% file: ub.tex
\section{Streaming Algorithms}
\label{sec:ub}
In this section, we present optimal approximation algorithms for $\mtwocsp$s using $O(\log n)$ space. In Theorem~\ref{thm:main} in Section~\ref{sec:theorem} we will prove that it is actually sufficient to design optimal algorithms for $\mcsp(\cG)$ in the following five cases $\cG\in\{\cspt, \cspor, \{\cspt, \cspor\}, \cspxor, \cspand\}$.
In Section~\ref{sec:alg trivial}, we present the trivial algorithm for \mtwocsp{}s, this algorithm turns our to be optimal for $\cG\in\{\cspt, \cspor, \cspxor\}$. Then we develop and analyze optimal algorithms for the cases $\cG=\cspand$ and $\cG=\{\cspt, \cspor\}$ in Sections~\ref{sec:alg and} and~\ref{sec:alg or}, respectively.

For ease of exposition, we will assume that input instances never contain unsatisfiable and tautological clauses (\textit{e.g.,} $(x\wedge\neg x), (x\vee \neg x)$). This assumption is without loss of generality, because a streaming algorithm can ignore unsatisfiable clauses and have a separate counter for tautological clauses.

\subsection{Trivial Algorithm}\label{sec:alg trivial}
First we present the trivial algorithm: this algorithm takes a \mtwocsp{} instance $\Psi$, counts the number of clauses $m=|\Psi|$ in it, and outputs the expected number of clauses satisfied by a uniform random assignment to the variables of $\Psi$. In Section~\ref{sec:lb} we will show that this algorithm gives the best streaming approximation not only in the case of \textsf{Max-2XOR} (the \textsf{Max-CUT} problem), but also in the case of \textsf{Max-2EOR}.

\begin{proposition}[Folklore]\label{thm:trivial alg}
For a function $f:\{0,1\}^2\rightarrow\{0,1\}$, let $\alpha_f\in[0,1]$ denote the fraction of~1s in its truth table.
Then for a set of allowed predicates $\cF$, we define $\alpha_\cF^{\text{tr}}=\min_{f\in\cF}\alpha_f$.
There exists a streaming algorithm that uses $O(\log n)$ space, and computes $\alpha_\cF^{\text{tr}}$-approximation for $\mcsp(\cF)$ with success probability~$1$.
\end{proposition}
For example, for the problem \meor{} (\emph{i.e.}, $\cF=\{\cspor\}$), we have $\alpha_{\cspor}=3/4$, as every clause is satisfied by 3 out of 4 possible assignments to its variables. Since the problem \mor{} (\emph{i.e.}, $\cF=\{\cspt,\cspor\}$) also allows clauses of length $1$ (which are satisfied by 1 out of 2 possible assignments to the variable), we have $\alpha_{\{\cspt,\cspor\}}=1/2$.
\begin{proof}[Proof of Proposition~\ref{thm:trivial alg}]
\mbox{}
\begin{algorithm}[ht] 
	\caption{$\alpha_\cF^{\text{tr}}$-approximation streaming algorithm for $\mcsp(\cF)$}\label{alg:trivial}
	\begin{algorithmic}[1]
	    \Input $\Psi$---an instance of $\mcsp(\cF)$.
		\State Use $O(\log n)$ bits to compute $m=|\Psi|$.
		\Output $v=\alpha_\cF^{\text{tr}}\cdot m$.
	\end{algorithmic}
\end{algorithm}

To prove that Algorithm~\ref{alg:trivial} computes an $\alpha_\cF^{\text{tr}}$-approximation, we need to show that (i) there exists an assignment $\sigma$ such that $\val_\Psi(\sigma)\geq v=\alpha_\cF^{\text{tr}}\cdot m$, and (ii) $v=\alpha_\cF^{\text{tr}}\cdot m\geq\alpha_\cF^{\text{tr}}\cdot\val_\Psi$. 

Note that since $\val_\Psi \leq |\Psi|=m$, (ii) holds trivially. 
The existence of an assignment $\sigma$ satisfying (i) is guaranteed by the following bound on the expected number of clauses satisfied by a uniform random assignment $\sigma$:
\[
\mathop{\mathbb{E}}_{\sigma}[\val_\Psi(\sigma)] = \sum_{C\in \Psi}\Pr_\sigma\left[C\text{ is satisfied by }\sigma\right]
= \sum_{C\in \Psi}\alpha_C\geq \alpha_\cF^{\text{tr}}\cdot m \, .
\]
\end{proof}

\begin{remark}
For an $(\alpha_\cF^{\text{tr}}-\epsilon)$-approximation, one can reduce the space usage of Algorithm~\ref{alg:trivial} to $O\left(\log\log{n}+\log(1/\epsilon)\right)$ bits by using the approximate counting algorithm of Morris~\cite{morris1978counting,gronemeier2009applying}.
\end{remark}

\begin{remark}
Formally, Algorithm~\ref{alg:trivial} only guarantees a $1/2$-approximation for the problem $\mcsp(\cspt)$, \emph{i.e.}, the problem where all clauses have length $1$. In this case, in order to achieve a $(1-\eps)$-approximation using  $O(\log{n})$ space for arbitrary constant $\eps>0$, one can use an $\ell_1$-sketch (Theorem~\ref{thm:l1sketch}) to approximate the bias vector of the input formula. Indeed, it is easy to see that for an instance $\Psi$ of $\mcsp(\cspt)$ with $m$ clauses, $\val_\Psi=(m+\bias(\Psi))/2$.
\end{remark}

We give $\alpha_\cG^{\text{tr}}$ for relevant sets of predicates in Table~\ref{table:trivial}.
\begin{table}[ht]
\centering
\def\arraystretch{1.5}
\begin{tabular}{|c|c|c|c|c|c|c|c|}
\hline
Type $\cG$       & $\cspt$ & $\cspor$     & $\{\cspt, \cspor\}$     & $\cspxor$      & $\cspand$     \\ \hline
$\alpha_\cG^{\text{tr}}$ & 1 & $\frac{3}{4}$ & $\frac{1}{2}$ & $\frac{1}{2}$ & $\frac{1}{4}$\\ \hline
$\alpha_\cG^{\text{opt}}$ & 1 & $\frac{3}{4}$ & $\frac{\sqrt{2}}{2}$ & $\frac{1}{2}$ & $\frac{4}{9}$ \\ \hline
\end{tabular}
\caption{For various sets of predicates $\cG$, the table presents (i) $\alpha_\cG^{\text{tr}}$---the approximation ratio guaranteed by the trivial algorithm for $\mcsp(\cG)$, and (ii) $\alpha_\cG^{\text{opt}}$---the optimal approximation ratio of streaming algorithms, proven in Sections~\ref{sec:ub} and~\ref{sec:lb} for $\mcsp(\cG)$. We have suppressed $(1-\eps)$ multiplicative factors for the case $\cG=\cspt$.}
\label{table:trivial}
\end{table}

As we show in the following sections, this trivial approximation algorithm can be improved for the $\mand$ and $\mor$ problems.

\subsection{Algorithm for  \textsf{Max-2AND} and  \textsf{Max-2EAND}}\label{sec:alg and}

Consider a \mand{} instance $\Psi'$ where all clauses are of length 1 or 2. Note that $\Psi'$ can be written as an equivalent \mand{} instance $\Psi$, where 1-clauses of $\Psi'$ are replaced with 2-clauses containing the same literal twice.\footnote{We only apply this transformation to \mand{} instances, because here it plays in our favor. For example, an \cspand{} clause with repeated literals is satisfied by a uniform random assignment with probability $1/2$, while an \cspand{} clause with distinct variables is satisfied with probability only $1/4$. For the case of \cspor{}, a clause with repeated literals would be satisfied only with probability $1/2$, while an \cspor{} clause with distinct variables would be satisfied with probability $3/4$.} In this section, we will consider such representation of every instance of \mand{}, \emph{i.e.}, we will assume that all clauses have exactly 2 (not necessarily distinct) literals. Note that in this case, the bias (see Definition~\ref{def:biasinstance}) of $\Psi$ is simply
\[
 \bias(\Psi) = \frac{1}{2}\sum_{i=1}^n |\posvar_i^{(2)}(\Psi) - \negvar_i^{(2)}(\Psi)| \, ,
\]
where $\posvar_i^{(2)}(\Psi)$ and $\negvar_i^{(2)}(\Psi)$ are the numbers of occurrences of $x_i$ and $\neg x_i$ in 2-clauses.

\cite{GVV17} gave lower and upper bounds for the maximum number of satisfied clauses $\val_\Psi$ in terms of $\bias(\Psi)$ and $m$ (the number of clauses in $\Psi$). For the sake of being self contained, and to verify that these bounds hold for our slightly more general case where 2-clauses may contain repeated literals, we present the proofs of these bounds in Lemma~\ref{lem:GVV17} in Section~\ref{sec:proof of and bias lower bound}.

\begin{restatable}[\cite{GVV17}]{lemma}{biasGVV}
\label{lem:GVV17}
Let $\Psi$ be a $\mand$ instance with $m$ clauses. Then
\[
\bias(\Psi)\leq \val_\Psi \leq\frac{m+\bias(\Psi)}{2} \,.
\]

\end{restatable}

We improve the lower bound of~\cite{GVV17} in the important regime of $\bias(\Psi)\leq m/3$ in the following lemma.

\begin{restatable}{lemma}{biaslb}
\label{lem:and bias lower bound}
Let $\Psi$ be a $\mand$ instance with $m$ clauses and $\bias(\Psi)\leq m/3$. Then
\[
\val_\Psi \geq
\frac{m}{4}+\frac{\bias(\Psi)^2}{4(m-2\bias(\Psi))} 
\geq 
\frac{2(m+\bias(\Psi))}{9}  \, .
\]
\end{restatable}
The proof of Lemma~\ref{lem:and bias lower bound} is based on \textit{biased random sampling}, and is postponed to Section~\ref{sec:proof of and bias lower bound}. For a pictorial view of this improvement, see Figure~\ref{fig:alg and}.

We are now ready to present a streaming algorithm that $(4/9)$-approximates \mand{} and \meand{}.

\begin{theorem}[$\frac{4}{9}$--approximation for $\mand$ and $\meand$]\label{thm:alg and}
For any $\epsilon\in(0,0.01)$, there exists a streaming algorithm that uses space $O(\eps^{-2}\log n)$ and computes $\left(\frac{4}{9}-\epsilon\right)$-approximation for $\mand$ and $\meand$ with success probability at least $3/4$.
\end{theorem}

\begin{proof}
The algorithm uses the bounds from Lemmas~\ref{lem:GVV17} and~\ref{lem:and bias lower bound} to approximate the value of a given instance of $\mand$.
\begin{algorithm}[ht]
	\caption{$\left(\frac{4}{9}-\epsilon\right)$-approximation streaming algorithm for $\mand$}\label{alg:maxand}
	\begin{algorithmic}[1]
		\Input $\Psi$---an instance of $\mand$. Error parameter $\epsilon\in(0,0.01)$.
			\State Approximate the $\ell_1$-norm of the bias vector with error $\delta=\epsilon/2$ (Theorem~\ref{thm:l1sketch}):
			\Statex Compute $B\in\left(1\pm \delta\right)\bias(\Psi)$.
			\State Count the number of clauses $m=|\Psi|$.
			\If{$B\in\left[0,\frac{m}{3} (1-\delta)\right]$}
				\Statex {\bf Output:} $v=\frac{2(m+B)}{9(1+\delta)}$.
			\Else
				\Statex {\bf Output:} $v=\frac{B}{(1+\delta)}$.
			\EndIf
	\end{algorithmic}
\end{algorithm}

To prove the correctness of Algorithm~\ref{alg:maxand}, we show that (i) $\val_\Psi\geq v$ and (ii) $v \geq \left( \frac{4}{9} - \epsilon \right)\cdot\val_\Psi$, where $v$ is the output of Algorithm~\ref{alg:maxand}.

\paragraph{(i) $\bm{v\leq\val_\Psi}$.}
Since $B$ is an $(1\pm\delta)$-approximation of the bias, with probability at least $3/4$ we have that $(1-\delta)\cdot\bias(\Psi)\leq B\leq(1+\delta)\cdot\bias(\Psi)$. 

 First, consider the case where $B\in\left[0,\frac{m}{3}(1-\delta)\right] \,:$
\[
v=\frac{2(m+B)}{9(1+\delta)}
\leq
\frac{2(1+\delta)(m+\bias(\Psi))}{9(1+\delta)}
=\frac{2(m+\bias(\Psi))}{9}
\leq \val_\Psi,
\]
where the last inequality uses the bound from Lemma~\ref{lem:and bias lower bound}.

Now consider the case where $B>\frac{m}{3}(1-\delta)$:
\[
v=\frac{B}{(1+\delta)}
\leq\bias(\Psi) 
\leq\val_\Psi,
\]
where the last inequality follows from the bound $\val_\Psi\geq\bias(\Psi)$ from Lemma~\ref{lem:GVV17}.

\paragraph{(ii) $\bm{v \geq \left( \frac{4}{9} - \epsilon \right)\cdot\val_\Psi}$.}
First, consider the case where $B\in\left[0,\frac{m}{3}(1-\delta)\right] \,
:$, 
\begin{align*}
    v=\frac{2(m+B)}{9(1+\delta)}
    \geq \frac{2(1-\delta)(m+\bias(\Psi))}{9(1+\delta)}
    \geq \frac{2(1-2\delta)(m+\bias(\Psi))}{9}
    \geq \left(\frac{4}{9}-\epsilon\right)\cdot\val_\Psi \, ,
\end{align*}
where the last inequality follows from the bound $\val_\Psi \leq\frac{m+\bias(\Psi)}{2}$ of Lemma~\ref{lem:GVV17} and $\delta=\epsilon/2$.

Now consider the case where $B > \frac{m}{3}(1-\delta)$. From Lemma~\ref{lem:GVV17}, $\val_\Psi\leq\frac{m+\bias(\Psi)}{2}$. Then

\begin{align*}
\frac{v}{\val_\Psi}
&\geq \frac{2v}{m+\bias(\Psi)}
= \frac{2B}{(1+\delta)(m+\bias(\Psi))}
\geq \frac{2B}{(1+\delta)(m+\frac{B}{1-\delta})}
\geq \frac{2B}{(1+3\delta)(m+B)}\\
&\geq \frac{\frac{2m(1-\delta)}{3}}{(1+3\delta)\frac{4m}{3}}
= \frac{1}{2}\cdot \frac{1-\delta}{1+3\delta}
>\frac{4}{9}
\end{align*}
for every $\delta<0.01$.

We conclude that Algorithm~\ref{alg:maxand} outputs a $(4/9-\epsilon)$-approximation for \textsf{Max-2AND} and \textsf{Max-2EAND}.
\end{proof}

\subsubsection{Proofs of Lemma~\ref{lem:GVV17} and Lemma~\ref{lem:and bias lower bound}}\label{sec:proof of and bias lower bound}
\biasGVV*
\begin{proof}

In order to prove the lower bound $\val_\Psi \geq\bias(\Psi)$, we give an assignment $\sigma$ to the input variables which satisfies at least $\bias(\Psi)$ clauses. This assignment $\sigma$ will greedily assign the value of each variable according to its bias: the variables which appear positively more often than negatively will be assigned~1, and the remaining variables will be assigned~0.

Recall that $\posvar_i^{(2)}(\Psi)$ and $\negvar_i^{(2)}(\Psi)$ denote the number of clauses where $x_i$ appears positively and negatively. For every variable $x_i$ with $\posvar_i^{(2)}(\Psi)\geq\negvar_i^{(2)}(\Psi)$, we set $\sigma(x_i)=1$, and we set $\sigma(x_i)=0$ otherwise. Note that the number of unsatisfied literals in this case is $\sum_i \min\left\{\posvar_i^{(2)}(\Psi), \negvar_i^{(2)}(\Psi)\right\}$. Thus, the number of unsatisfied clauses is also bounded from above by $\sum_i \min\left\{\posvar_i^{(2)}(\Psi), \negvar_i^{(2)}(\Psi)\right\}$.

From 
\begin{align*}
2\bias(\Psi)=&\sum_i \max\left\{\posvar_i^{(2)}(\Psi), \negvar_i^{(2)}(\Psi)\right\}-\min\left\{\posvar_i^{(2)}(\Psi), \negvar_i^{(2)}(\Psi)\right\}\\
2m=&\sum_i \max\left\{\posvar_i^{(2)}(\Psi), \negvar_i^{(2)}(\Psi)\right\}+\min\left\{\posvar_i^{(2)}(\Psi), \negvar_i^{(2)}(\Psi)\right\}
\end{align*}
we have that
\begin{align}
\label{eq:and_bound_aux}
\sum_i \min\left\{\posvar_i^{(2)}(\Psi), \negvar_i^{(2)}(\Psi)\right\}=m-\bias(\Psi) \, .
\end{align}
Thus, 
\[
\val_\Psi(\sigma) \geq m - (m-\bias(\Psi)) = \bias(\Psi) \, .
\]

For the upper bound of $\val_\Psi \leq\frac{m}{2}+\frac{\bias(\Psi)}{2}$ we note that for every assignment $\sigma$, the number of unsatisfied literals is at least $\sum_i \min\left\{\posvar_i^{(2)}(\Psi), \negvar_i^{(2)}(\Psi)\right\}$. (Since $x_i=1$ produces $\posvar_i^{(2)}(\Psi)$ unsatisfied literals, while $x_i=0$ produces $\negvar_i^{(2)}(\Psi)$ unsatisfied literals.)
Thus, the number of unsatisfied clauses is at least $\frac{1}{2}\sum_i \min\left\{\posvar_i^{(2)}(\Psi), \negvar_i^{(2)}(\Psi)\right\}$. From~\eqref{eq:and_bound_aux}, we have that for every assignment $\sigma$,
\[
\val_\Psi(\sigma) \leq m - \frac{1}{2}(m-\bias(\Psi)) = \frac{m+\bias(\Psi)}{2} \, .
\]

\end{proof}
\biaslb*
\begin{proof}
First we show that for every $\mand$ instance $\Psi$ with $m$ clauses and $\bias(\Psi)\leq m/3$, there exists an assignment 
$\sigma$ s.t. 
\[\val_\Psi(\sigma)\geq \frac{m}{4}+\frac{\bias(\Psi)^2}{4(m-2\bias(\Psi))} \ .\]

Without loss of generality we can assume that every variable appears in $\Psi$ positively at least as many times as it appears negatively, \emph{i.e.}, $\posvar_i^{(2)}(\Psi)\geq\negvar_i^{(2)}(\Psi)$ for every $i\in[n]$.\footnote{Indeed, given a instance $\Psi$ where $\posvar_i^{(2)}(\Psi)<\negvar_i^{(2)}(\Psi)$, we can consider the instance $\Psi'$ where every $x_i$ is replaced with $\neg x_i$, and vice versa. We have that $\posvar_i^{(2)}(\Psi')\geq\negvar_i^{(2)}(\Psi')$, $\bias(\Psi)=\bias(\Psi')$, and every assignment for $\Psi'$ is uniquely mapped to the corresponding assignment for $\Psi$ satisfying the same number of clauses.} We prove the existence of such an assignment $\sigma$ by giving a distribution of assignments whose expected number of satisfied clauses is at least $\frac{2(m+\bias(\Psi))}{9}$. 
Let $\gamma\in[0,0.5]$ be a parameter to be assigned later. For each variable $x_i$, we assign $x_i = 1$ with probability $\frac{1}{2}+\gamma$, and $x_i = 0$ with probability $\frac{1}{2}-\gamma$.\footnote{Note that if we set $\gamma=0$, the algorithm becomes the trivial random sampling, and if we set $\gamma=0.5$, the algorithm becomes the greedy algorithm from~\cite{GVV17}.} Let $k_0$, $k_1,$ and $k_2$ denote the number of clauses with zero, one, and two positive literals. Observe that $m=k_0+k_1+k_2$ and
\begin{align*}
2\bias(\Psi) &= \sum_{i\in[n]}|\posvar_i^{(2)}(\Psi)-\negvar_i^{(2)}(\Psi)|
=(2k_2+k_1)-(k_1+2k_0)=2(k_2-k_0) \, .
\end{align*}

Let us now compute the expected number of satisfied $\cspand$ clauses under the biased distribution described above. Note that a clause with two (not necessarily distinct) positive literals is satisfied with probability at least $\min\left\{\left(\frac{1}{2}+\gamma\right)^2, \frac{1}{2}+\gamma\right\}=\left(\frac{1}{2}+\gamma\right)^2$. Similarly, a clause with two negative literals is satisfied with probability at least $\left(\frac{1}{2}-\gamma\right)^2$, and a clause with a positive and negative literals (corresponding to different variables) is satisfied with probability $\left(\frac{1}{2}-\gamma\right)\left(\frac{1}{2}+\gamma\right)$.

\begin{align*}
\Exp_{\sigma}\left[\val_\Psi(\sigma)\right]&= \sum_{i=0}^2k_i\cdot\Pr_\sigma\left[\text{a clause with $i$ positive literals is satisfied by } \sigma\right]\\
&=k_0\cdot\left(\frac{1}{2}-\gamma\right)^2+k_1\cdot\left(\frac{1}{2}-\gamma\right)\left(\frac{1}{2}+\gamma\right)+k_2\cdot\left(\frac{1}{2}+\gamma\right)^2\\
&= \frac{k_0+k_1+k_2}{4}+(k_2 - k_0)\cdot\gamma+(k_2-k_1+k_0)\cdot\gamma^2 \\
&=\frac{m}{4} + \bias(\Psi)\cdot\gamma+(2(k_2+k_0)-m)\cdot\gamma^2\\
&\geq \frac{m}{4} + \bias(\Psi)\cdot\gamma+(2\bias(\Psi)-m)\cdot\gamma^2 \, ,
\end{align*}
where we used that $m=k_0+k_1+k_2$ and $2\bias(\Psi)=2(k_2-k_0)\leq2(k_2+k_0)$.

Since $\bias(\Psi)\in[0,m/3]$, we can set $\gamma=\frac{\bias(\Psi)}{2(m-2\bias(\Psi))}\in[0,0.5]$ and have that
\begin{align*}
\Exp_{\sigma}\left[\val_\Psi(\sigma)\right]\geq\frac{m}{4}+\frac{\bias(\Psi)^2}{4(m-2\bias(\Psi))} \, .
\end{align*}

Finally, it remains to show that $\frac{m}{4}+\frac{\bias(\Psi)^2}{4(m-2\bias(\Psi))} \geq \frac{2(m+\bias(\Psi))}{9}\, :$
\begin{align*}
\frac{m}{4}+\frac{\bias(\Psi)^2}{4(m-2\bias(\Psi))} 
&=\frac{2(m+\bias(\Psi))}{9} + \frac{m-8\bias(\Psi)}{36} + \frac{\bias(\Psi)^2}{4(m-2\bias(\Psi))}\\
&=\frac{2(m+\bias(\Psi))}{9} + \frac{(m-8\bias(\Psi))(m-2\bias(\Psi))+9\bias(\Psi)^2}{36(m-2\bias(\Psi))}\\
&=\frac{2(m+\bias(\Psi))}{9} +\frac{(5\bias(\Psi)-m)^2}{36(m-2\bias(\Psi))}\\
&\geq \frac{2(m+\bias(\Psi))}{9} \, ,
\end{align*}
which holds for every $\bias(\Psi)\in[0,m/3]$.
\end{proof}

\subsection{Algorithm for  \textsf{Max-2OR}}\label{sec:alg or}
For the case of \textsf{Max-2OR}, it is crucial to distinguish 1- and 2-clauses. Therefore, we treat clauses containing two identical literals as 1-clauses. We denote the number of 1-clauses of $\Psi$ by $m_1$, and the number of 2-clauses by $m_2$. In particular, the total number of clauses is $m=m_1+m_2$.

In Lemmas~\ref{lem:ub2} and~\ref{lem:or bias lower bound} we give  upper and lower bounds on $\val_\Psi$ in terms of $m_1, m_2$, and $\bias(\Psi)$, we postpone their proofs to Section~\ref{sec:proofs_or}. In this section we prove that the ratio between the presented lower and upper bounds is bounded by $\frac{\sqrt{2}}{2}$, and that there is a $O(\log{n})$-space algorithm that sketches the lower bounds of Lemma~\ref{lem:or bias lower bound}  on $\val_\Psi$.

When the $\bias$ of $\Psi$ is large (say, $\bias(\Psi)=m$), it might be possible to satisfy all $m$ clauses of $\Psi$, so no non-trivial upper bounds on $\val_\Psi$ can be proven in terms of $\bias$ in this case. Even if the $\bias$ is low (say, $\bias(\Psi)=0$), but the formula does not contain 1-clauses, it might still be possible to satisfy all clauses of $\Psi$. (\emph{E.g.}, if all clauses of $\Psi$ contain one positive and one negative literal.) It turns out that for the optimal approximation ratio, we need to bound from above $\val_\Psi$ in the case of low $\bias$ and large number of 1-clauses. 
\begin{restatable}{lemma}{ubtwo}
\label{lem:ub2}
Let $\Psi$ be a $\mor$ instance with $m_1$ 1-clauses, and $m_2$ 2-clauses. Then
\[
\val(\Psi) \leq \min\left\{m_1+m_2, \frac{m_1+2m_2+\bias(\Psi)}{2}\right\} 
\, .
\]
\end{restatable}

The trivial algorithm guarantees that for every $\mor$ instance $\Psi$, $\val_\Psi\geq m_1/2+3m_2/4$. While this bound is tight in terms of $m_1$ and $m_2$, for instances with high $\bias>m_2/2$, we prove a better lower bound of $\val \geq (m_1+m_2+\bias(\Psi))/2$. Clearly, this bound is not sufficient for a better than $1/2$-approximation in the case of low $\bias(\Psi)=0$. In order to handle this case, we design a distribution of assignments which in expectation satisfy a large number of clauses in formulas with low $\bias$.
\begin{restatable}{lemma}{biaslbtwo}
\label{lem:or bias lower bound}
Let $\Psi$ be a $\mor$ instance with $m_1$ 1-clauses, and $m_2$ 2-clauses. Then
\begin{enumerate}[leftmargin=*]
    \item \(\val_\Psi \geq \frac{m_1+m_2+\bias(\Psi)}{2} \, ;\)
    \item if $\bias(\Psi)\leq m_2$, then
    \[\val_\Psi \geq
\frac{m_1}{2}+\frac{3m_2}{4}+\frac{\bias(\Psi)^2}{4m_2}  \, .\]
\end{enumerate}
\end{restatable}

We will also use the following simple claim.
\begin{claim}
\label{claim}
For every $x\geq 0, y>0$:
\[
\frac{2x+3y+x^2/y}{4(x+y)}\geq\frac{\sqrt{2}}{2}
\, .
\]
\end{claim}
\begin{proof}
Let $z=\frac{x}{y}+1$, then
\begin{align*}
    \frac{2x+3y+x^2/y}{4(x+y)}
    =\frac{2\frac{x}{y}+3+\frac{x^2}{y^2}}{4\left(\frac{x}{y}+1\right)}
    =\frac{z^2+2}{4z}
    =\frac{z}{4}+\frac{1}{2z}
    \geq\frac{\sqrt{2}}{2}
    \, ,
\end{align*}
by the inequality of arithmetic and geometric means.
\end{proof}

Now we are ready to present an approximation algorithm for the \textsf{Max-2OR} problem.
\begin{theorem}[$\frac{\sqrt{2}}{2}$--approximation for $\mor$]\label{thm:alg or}
For any $\epsilon\in(0,0.01)$, there exists a streaming algorithm that uses space $O(\eps^{-2}\log n)$ and computes $\left(\frac{\sqrt{2}}{2}-\epsilon\right)$-approximation for $\mor$ with success probability at least~$3/4$.
\end{theorem}
\begin{proof}
We prove that Algorithm~\ref{alg:maxor} computes a $\left(\frac{\sqrt{2}}{2}-\epsilon\right)$-approximation by showing that (i) $v\leq \val_\Psi$, and (ii) $v \geq \left( \frac{\sqrt{2}}{2} - \epsilon \right)\cdot\val_\Psi$, where $v$ is the output of the algorithm. Recall that by the guarantee of Theorem~\ref{thm:l1sketch}, with probability at least $3/4$:
\[
(1-\delta)\bias(\Psi) \leq B \leq (1+\delta)\bias(\Psi)
\, .
\]
\begin{algorithm}[ht]
	\caption{$\left(\frac{\sqrt{2}}{2}-\epsilon\right)$-approximation streaming algorithm for $\mor$}\label{alg:maxor}
	\begin{algorithmic}[1]
		\Input $\Psi$---an instance of $\mor$. Error parameter $\epsilon\in(0,0.01)$.
			\State Approximate the $\ell_1$-norm of the bias vector with error $\delta=\epsilon/4$ (Theorem~\ref{thm:l1sketch}):
			\Statex Compute $B\in\left(1\pm \delta\right)\bias(\Psi)$.
			\State Count the number of 1- and 2- clauses $m_1$ and $m_2$.
			\If{$B\in\left[0,(1-\delta)m_2\right]$}
				\Statex {\bf Output:} $v=\frac{(1-\delta)^2(2m_1+3m_2+B^2/m_2)}{4}$.
			\Else
				\Statex {\bf Output:} $v=\frac{(1-\delta)(m_1+m_2+B)}{2}$.
			\EndIf
	\end{algorithmic}
\end{algorithm}

\paragraph{(i) $\bm{v\leq \val_\Psi}$.}
First, note that $(1-\delta)B\leq(1-\delta)(1+\delta)\bias(\Psi)\leq\bias(\Psi)$.
Next, if $B\leq(1-\delta)m_2$, then $\bias(\Psi)\leq B/(1-\delta)\leq m_2$, and, thus, $\val_\Psi \geq
\frac{m_1}{2}+\frac{3m_2}{4}+\frac{\bias(\Psi)^2}{4m_2}$ by the second bound in Lemma~\ref{lem:or bias lower bound}. Then
\[
v=\frac{(1-\delta)^2(2m_1+3m_2+B^2/m_2)}{4}
\leq \frac{(2m_1+3m_2+\bias(\Psi)^2/m_2)}{4}
\leq \val_\Psi
\, .
\]

If $B>(1-\delta)m_2$, then 
\[
v=\frac{(1-\delta)(m_1+m_2+B)}{2}
\leq \frac{m_1+m_2+\bias(\Psi)}{2}
\leq \val_\Psi
\]
by the first bound in Lemma~\ref{lem:or bias lower bound}.

\paragraph{(ii) $\bm{v \geq \left( \frac{\sqrt{2}}{2} - \epsilon \right)\cdot\val_\Psi}$.}

Let us consider three cases.
\begin{enumerate}
\item $B\leq(1-\delta)m_2$ and $m_1\leq \bias(\Psi)$.\\
    In this case the output of the algorithm is 
    \begin{align*}
    v&=\frac{(1-\delta)^2(2m_1+3m_2+B^2/m_2)}{4}\\
    &\geq \frac{(1-\delta)^4(2m_1+3m_2+\bias(\Psi)^2/m_2)}{4}\\ 
    &\geq \frac{(1-4\delta)(2m_1+3m_2+\bias(\Psi)^2/m_2)}{4}
    \, .
    \end{align*}
   
    From the upper bound $\val_\Psi\leq m_1+m_2$ of Lemma~\ref{lem:ub2}, we have that
\begin{align*}
\frac{v}{\val_\Psi}
&\geq (1-4\delta)\cdot\frac{2m_1+3m_2+\bias(\Psi)^2/m_2}{4(m_1+m_2)}\\
&\geq (1-4\delta)\cdot\frac{2\bias(\Psi)+3m_2+\bias(\Psi)^2/m_2}{4(\bias(\Psi)+m_2)}\\
&\geq(1-4\delta)\cdot\frac{\sqrt{2}}{2} 
=(1-\epsilon)\cdot\frac{\sqrt{2}}{2} 
\, ,
\end{align*}
where the second inequality follows from $m_1\leq\bias(\Psi)$, and the last inequality follows from Claim~\ref{claim}.

\item $B\leq(1-\delta)m_2$ and $m_1>\bias(\Psi)$.\\
From the upper bound $\val_\Psi\leq \frac{m_1+2m_2+\bias(\Psi)}{2}$ of Lemma~\ref{lem:ub2}:
\begin{align*}
\frac{v}{\val_\Psi}
&\geq (1-4\delta)\cdot\frac{2m_1+3m_2+\bias(\Psi)^2/m_2}{2(m_1+2m_2+\bias(\Psi))}\\
&\geq (1-4\delta)\cdot\frac{2\bias(\Psi)+3m_2+\bias(\Psi)^2/m_2}{4(\bias(\Psi)+m_2)}\\
&\geq(1-4\delta)\cdot\frac{\sqrt{2}}{2}
=(1-\epsilon)\cdot\frac{\sqrt{2}}{2}
\, ,
\end{align*}
where the second inequality is due to $m_1>\bias(\Psi)$, and the last inequality is due to Claim~\ref{claim}.

\item $B>(1-\delta)m_2$.\\
From the bound in~Lemma~\ref{lem:ub2}: 
\begin{align*}
\val_\Psi &\leq \min\left\{m_1+m_2,\frac{m_1+2m_2+\bias(\Psi)}{2}\right\}\\
&\leq \frac{1}{3}\cdot(m_1+m_2)+\frac{2}{3}\cdot\frac{m_1+2m_2+\bias(\Psi)}{2}\\
&=\frac{2m_1+3m_2+\bias(\Psi)}{3}\\
&\leq \frac{2m_1+3m_2+B}{3(1-\delta)}
\, .
\end{align*}
In this case, the output of the algorithm is
$v=\frac{(1-\delta)(m_1+m_2+B)}{2}$. Then
\[
\frac{v}{\val_\Psi}
    \geq \frac{3(1-\delta)^2}{2}\cdot
    \frac{m_1+m_2+B}{2m_1+3m_2+B}
    \geq \frac{3(1-\delta)^2}{2}\cdot
    \frac{m_1+m_2(2-\delta)}{2m_1+4m_2}
    \geq\frac{3(1-\delta)^2(2-\delta)}{8}
    \geq\frac{\sqrt{2}}{2} \, ,
\]
where the second inequality is due to $B>(1-\delta)m_2$, and the last one holds for every $\delta<0.01$.
\end{enumerate}
\end{proof}

\subsubsection{Proofs of Lemma~\ref{lem:ub2} and Lemma~\ref{lem:or bias lower bound}}\label{sec:proofs_or}

\ubtwo*
\begin{proof}
Since $m_1+m_2$ is the number of clauses in~$\Psi$, the first bound $\val(\Psi) \leq m_1+m_2$ holds trivially.

First we negate all variables of $\Psi$ with $\bias_i(\Psi)<0$. This transformation does not change $\bias(\Psi), m_1, m_2, \val(\Psi)$, and every assignment of the variables of the original instance can be uniquely mapped to a corresponding assignment for the new instance satisfying the same number of clauses.
Therefore, without loss of generality, for every $i\in[n]$,
\[
\posvar_i^{(1)}(\Psi)+\frac{\posvar_i^{(2)}(\Psi)}{2}-\negvar_i^{(1)}(\Psi)-\frac{\negvar_i^{(2)}(\Psi)}{2}\geq0
\, .
\]

Consider an assignment $\sigma$ to the variables of $\Psi$. We need to show that $\val_\Psi(\sigma)\leq\frac{m_1+2m_2+\bias(\Psi)}{2}$. Let $T$ be the set of (indices of) variables of $\sigma$ assigned the value $1$. Then the number of $1$ clauses satisfied by $\sigma$ is
\[
S_1=\sum_{i\in T}\posvar_i^{(1)}(\Psi)+\sum_{i\not\in T}\negvar_i^{(1)}(\Psi)
\, .
\]

Let $S_2$ denote the number of 2-clauses satisfied by $\sigma$. We will show that
\begin{align}
\label{eq:s2}
S_2 \leq \min\left\{m_2, \bias(\Psi)+m_1+m_2-2S_1\right\} \, .
\end{align}
First we show how~\eqref{eq:s2} finishes the proof of the lemma, and then prove~\eqref{eq:s2}.

Indeed, then the number of clauses satisfied by $\sigma$ is bounded from above by
\begin{align*}
\val_\Psi(\sigma) 
&\leq S_1+S_2 \\
&\leq S_1 + \min\left\{m_2, \bias(\Psi)+m_1+m_2-2S_1\right\}\\
&\leq S_1 + \frac{m_2}{2}+\frac{\bias(\Psi)+m_1+m_2-2S_1}{2}\\
&=\frac{m_1+2m_2+\bias(\Psi)}{2}
\, .
\end{align*}

Now we will prove the bound~\eqref{eq:s2}. The bound $S_2\leq m_2$ is trivial, since $m_2$ is the total number of 2-clauses in the instance. The number of 2-clauses satisfied by variables set to $1$ is bounded from above by $\sum_{i\in T}\posvar_i^{(2)}(\Psi)$, and the number of 2-clauses satisfied by variables set to $0$ is bounded by $\sum_{i\not\in T}\negvar_i^{(2)}(\Psi)$. Therefore, 
\begin{align}
    S_2\leq \sum_{i \in T}\posvar_i^{(2)}(\Psi)+\sum_{i\not\in T}\negvar_i^{(2)}(\Psi) \, .
    \label{eq:final}
\end{align}
Recall that
\begin{align}
\bias(\Psi)
&= \sum_{i\in[n]} \posvar_i^{(1)}(\Psi)+\frac{\posvar_i^{(2)}(\Psi)}{2}-\negvar_i^{(1)}(\Psi)-\frac{\negvar_i^{(2)}(\Psi)}{2}\label{eq1}\, ,\\
m_1&=\sum_{i\in[n]}\posvar_i^{(1)}(\Psi)+\negvar_i^{(1)}(\Psi)\label{eq2}\, ,\\
m_2&=\sum_{i\in[n]}\frac{\posvar_i^{(2)}(\Psi)}{2}+\frac{\negvar_i^{(2)}(\Psi)}{2}\label{eq3}\, ,\\
-2S_1&=-2\sum_{i\in T}\posvar_i^{(1)}(\Psi)-2\sum_{i\not\in T}\negvar_i^{(1)}(\Psi)\label{eq4}\, ,
\end{align}
and since $\bias_i(\Psi)\geq 0$ for every $i$:
\begin{align}
    0\geq -2\sum_{i\not\in T}\bias_i(\Psi)=-\sum_{i\not\in T}2\posvar_i^{(1)}(\Psi)+\posvar_i^{(2)}(\Psi)-2\negvar_i^{(1)}(\Psi)-\negvar_i^{(2)}(\Psi) \label{eq5}\, .
\end{align}

Summing \eqref{eq1}, \eqref{eq2}, \eqref{eq3}, \eqref{eq4}, and \eqref{eq5} gives
\[
\bias(\Psi)+m_1+m_2-2S_1\geq \sum_{i\in T} \posvar_i^{(2)}(\Psi)+\sum_{i\not\in T} \negvar_i^{(2)}(\Psi)\geq S_2,
\]
where the last inequality uses~\eqref{eq:final}. This finishes the proof of~\eqref{eq:s2} and the proof of the lemma.
\end{proof}

\biaslbtwo*
\begin{proof}
Without loss of generality, we assume that for every $i\in[n], \bias_i(\Psi)\geq 0$. (Again, we can negate all variables with $\bias_i(\Psi)<0$, and define a bijection between the assignments for the two formulas.)
Therefore, for every $i\in[n],$
\[
\posvar_i^{(1)}(\Psi)+\frac{\posvar_i^{(2)}(\Psi)}{2}-\negvar_i^{(1)}(\Psi)-\frac{\negvar_i^{(2)}(\Psi)}{2}\geq0
\, .
\]
Let $p_1$ and $n_1$ be the numbers of 1-clauses with positive and negative literals in $\Psi$. Let $k_0, k_1$, and $k_2$ denote the numbers of 2-clauses with 0, 1, and 2 positive literals. Then $m_1=p_1+n_1$, and $m_2=k_0+k_1+k_2$.

Note that
\begin{align*}
\bias(\Psi) 
= \sum_i \posvar_i^{(1)}(\Psi)+\frac{\posvar_i^{(2)}(\Psi)}{2}-\negvar_i^{(1)}(\Psi)-\frac{\negvar_i^{(2)}(\Psi)}{2}
= p_1-n_1+k_2-k_0 
\, .
\end{align*}

Consider the distribution of assignments to the variables of $\Psi$, where every variable $x_i$ is assigned the value 1 independently with probability $(\frac{1}{2}+\gamma)$, for a parameter $\gamma\in[0, 0.5]$ to be assigned later. The expected number of satisfied 1-clauses under this distribution is
\[
S_1 = \sum_i \left(\frac{1}{2}+\gamma\right)\cdot\posvar_i^{(1)}(\Psi)+\left(\frac{1}{2}-\gamma\right)\cdot\negvar_i^{(1)}(\Psi) = \left(\frac{1}{2}+\gamma\right)p_1 + \left(\frac{1}{2}-\gamma\right)n_1 = \frac{m_1}{2}+\gamma(p_1-n_1) 
\, .
\]

Since every 2-clause contains distinct variables, the expected number of satisfied 2-clauses is
\begin{align*}
S_2 &= k_0\cdot \left(1-\left(\frac{1}{2}+\gamma\right)^2 \right)
+ k_1\cdot \left(1-\left(\frac{1}{2}+\gamma\right)\left(\frac{1}{2}-\gamma\right) \right)
+ k_2\cdot \left(1-\left(\frac{1}{2}-\gamma\right)^2 \right)\\
&= \frac{3m_2}{4} + \gamma\cdot\left(k_2-k_0\right)-\gamma^2\cdot\left(k_0+k_2-k_1\right)\\
&\geq \frac{3m_2}{4} + \gamma\cdot\left(k_2-k_0\right)-m_2\gamma^2
\, .
\end{align*}
Let us now compute the expected number of clauses satisfied by an assignment $\sigma$ from the distribution defined above.

\begin{align*}
\Exp_{\sigma}\left[\val_\Psi(\sigma)\right] 
= S_1 + S_2
&\geq \frac{m_1}{2}+\gamma(p_1-n_1) + \frac{3m_2}{4} + \gamma\cdot\left(k_2-k_0\right)-2m_2\gamma^2\\
&=\frac{m_1}{2}+ \frac{3m_2}{4} + \gamma\bias(\Psi) - m_2\gamma^2
\, .
\end{align*}
First, we set $\gamma=\frac{1}{2}$ and derive the first bound:
\[
\val_\Psi\geq\Exp_{\sigma}\left[\val_\Psi(\sigma)\right] \geq\frac{m_1+m_2+\bias(\Psi)}{2} \, .
\]
Now, for the case where $\bias(\Psi)\leq m_2$, we set
$\gamma=\frac{\bias(\Psi)}{2m_2}\in[0,0.5]$, and derive the second bound:
\[
\val_\Psi\geq\Exp_{\sigma}\left[\val_\Psi(\sigma)\right] \geq\frac{m_1}{2}+\frac{3m_2}{4}+\frac{\bias(\Psi)^2}{4m_2}  \, .
\]
\end{proof}

%% file: lb.tex
\section{Space Lower Bounds for Approximating Boolean \textsf{Max-2CSP}}
\label{sec:lb}

In this section, we establish space lower bounds for streaming approximations for all \mtwocsp{}s. In Theorem~\ref{thm:main} in Section~\ref{sec:theorem} we will show that it suffices to prove lower bounds for $\mcsp(\cG)$ for the following four cases $\cG\in\{\cspor, \{\cspt, \cspor\}, \cspxor, \cspand\}$. A linear space lower bound for the case $\cG=\cspxor$ is proven by Kapralov and Krachun~\cite{KK19}. We use this result to prove a linear lower bound for the case $\cF=\cspor$ in Section~\ref{sec:xor to or}. We prove the two remaining lower bounds by reductions from the communication complexity problem \DBHP~\cite{KKS15}. In Section~\ref{sec:dbhp}, we present a general framework for proving such lower bounds, while in Sections~\ref{sec:reduction and} and~\ref{sec:reduction or} we give specific reductions for the \mand{} and \mor{} problems. Finally, Sections~\ref{sec:proof of dbhp} and~\ref{sec:proof of dbhp thm} contain the proofs of some technical results used in the framework in Section~\ref{sec:dbhp}.

\subsection{From \mexor~to \meor}\label{sec:xor to or}
In this section, we give a simple streaming reduction from \mcut{} to \meor{}, which asserts that a better than trivial $3/4$-approximation for \meor{} would lead to a better then trivial $1/2$-approximation for \mcut{}. Since the latter is known to require linear space~\cite{KK19}, we get a linear lower bound on the space complexity of $(3/4+\eps)$-approximations of \meor{}.
\begin{lemma}[Folklore]\label{lem:exor to eor}
Let $\Psi_\cspxor$ be a $\mexor$ instance with $m$ clauses. Consider the following reduction from $\Psi_\cspxor$ to $\Psi_\cspor$, a $\meor$ instance: For every clause $(x \oplus y)$ in $\Psi_\cspxor$, we add clauses $(x \vee y)$ and $(\neg x \vee \neg y)$ to $\Psi_\cspor$. Then
\[
\val_{\Psi_\cspor} = m + \val_{\Psi_\cspxor}
\, .
\]
\end{lemma}
\begin{proof}
It suffices to show that for every assignment $\sigma$, $\val_{\Psi_\cspor}(\sigma) = m + \val_{\Psi_\cspxor}(\sigma)$. Suppose $\sigma$ satisfies the clause $x\oplus y$ in $\Psi_\cspxor$, then $\sigma(x)\neq \sigma(y)$. In this case, $\sigma$ satisfies both the corresponding clauses, $(x\vee y)$ and $(\neg x \vee \neg y)$ in $\Psi_\cspor$. On the other hand, if $\sigma$ does not satisfy $x\oplus y$ in $\Psi_\cspxor$, then $\sigma(x)=\sigma(y)$. In this case, $\sigma$ satisfies exactly one of the corresponding clauses in $\Psi_\cspor$.

\end{proof}
\begin{corollary}\label{cor:eor lb}
For any constant $\epsilon>0$, any streaming algorithm that $(3/4+\epsilon)$-approximates $\meor$ with success probability at least $3/4$ requires $\Omega(n)$ space.
\end{corollary}
\begin{proof}
Let $\ALG$ be a $(3/4+\epsilon)$-approximate algorithm for \meor{}. We will show that there exists a streaming algorithm of the same space complexity as $\ALG$ which $(1/2+4\eps/3)$-approximates \mexor{}. This, together with the $\Omega(n)$ space lower bound for $(1/2+\eps)$-approximations for \mexor{} (Theorem~\ref{thm:kk19}), will finish the proof.

Given a \mexor{} instance $\Psi_\cspxor$ with $m$ clauses, we use Lemma~\ref{lem:exor to eor} to convert it into a \meor{} instance $\Psi_\cspor$. Let $v$ be the output of the algorithm $\ALG$ on $\Psi_\cspor$, then we output $\max\{m/2, v-m\}$ as an approximation to $\val_{\Psi_\cspxor}$. It remains to show that 
$\left(\frac{1}{2}+\frac{4\epsilon}{3}\right)\cdot\val_{\Psi_\cspxor}
\leq
\max\{m/2, v-m\} 
\leq \val_{\Psi_\cspxor}$.

First, by Lemma~\ref{lem:exor to eor}
\[
v-m\leq\val_{\Psi_\cspor}-m=\val_{\Psi_\cspxor}.
\]
Together with the trivial bound $\val_{\Psi_\cspxor}\geq m/2$, this establishes that $\max\{m/2,v-m\}\leq\val_{\Psi_\cspxor}$.
Second,
\begin{align*}
\max\left\{\frac{m}{2},v-m\right\}&\geq\frac{1}{3}\cdot\frac{m}{2}+\frac{2}{3}\cdot(v-m)=\frac{m}{6}+\frac{2}{3}\cdot\left(\frac{3}{4}+\epsilon\right)\cdot\val_{\Psi_\cspor}-\frac{2m}{3}\\
&=\left(\frac{1}{2}+\frac{2\epsilon}{3}\right)\cdot(\val_{\Psi_\cspxor}+m)-\frac{m}{2}\\
&=\left(\frac{1}{2}+\frac{2\epsilon}{3}\right)\cdot\val_{\Psi_\cspxor}+\frac{2\epsilon m}{3}\\
&\geq\left(\frac{1}{2}+\frac{4\epsilon}{3}\right)\cdot\val_{\Psi_\cspxor} \, .
\end{align*}
\end{proof}

\subsection{Distributional Boolean Hidden Partition (\DBHP) Problem}\label{sec:dbhp}
We prove lower bounds for \meand{} and \mor{} in two steps. Recall that the goal of the players in $\DBHP$ is to distinguish between two distributions $\yes$ and $\no$. First, we show a reduction from $\DBHP$ to $\mcsp(\cG)$. This induces a $\yes$ and a $\no$ distributions of instances of $\mcsp(\cG)$, corresponding to the $\yes$ and $\no$ cases of $\DBHP$. Next, we show that with high probability there is a gap between the optimal value of instances from the $\yes$ and $\no$ distributions. The ratio $\alpha$ between these optimal values will be the upper bound on the approximation ratio of space-efficient streaming algorithms. Informally,  any $(\alpha+\epsilon)$-approximate streaming algorithm with space $s$ distinguishes the distributions $\yes$ and $\no$, and, therefore, can be converted into a communication protocol for $\DBHP$ that uses $s$ bits of communication. Since Kapralov, Khanna, and Sudan~\cite{KKS15} proved that any communication protocol for $\DBHP$ requires at least $\Omega(\sqrt{n})$ bits of communication, the corresponding space lower bound for streaming algorithms follows.

Before presenting the framework for streaming lower bounds, we will need to define \DBHP{} and slightly adjust it to our setting.

For $n\in\N$ and $p\in[0,1]$, by $G(n,p)$ we denote the Erd\"{o}s-R\'{e}nyi distribution of undirected graphs with $n$ vertices, where each edge is chosen independently with probability $p$.

\begin{definition}[$\DBHP$]\label{def:dbhp}
Let $n\in\N$, $\beta\in(0,1/16)$ be parameters. Let $X^*\in\{0,1\}^n$ be a uniformly random vector, and $G$ be a random graph sampled from $G(n,2\beta/n)$. Let $r$ be the number of edges in $G$, and ${M\in\{0,1\}^{r\times n}}$ be the edge-vertex incidence matrix of~$G$. We will consider the following three distributions of a vector $w\in\{0,1\}^r$.
\begin{itemize}
\item ($\yes$ distribution) $w=MX^*\in\{0,1\}^r$, where the arithmetic is over $\mathbb{F}_2$;
\item ($\no$ distribution) $w=\mathbf{1}+MX^*\in\{0,1\}^r$, where $\mathbf{1}\in\mathbb{F}_2^r$ is the all 1s vector, and the arithmetic is over $\mathbb{F}_2^r$;
\item ($\ono$ distribution) $w$ be uniformly sampled from $\{0,1\}^r$.
\end{itemize}

For a pair of distinct distributions $\cD\neq \cD'\in\{\yes,\no,\ono\}$, we consider the following decisional 2-player one-way communication problem $\DBHP_{\cD,\cD'}(n,\beta)$. Alice receives $X^*\in\{0,1\}^n$, and Bob receives $(M,w)$ as their private inputs, where $w$ is sampled from $\cD$ or $\cD'$ with probability $1/2$. A communication protocol $\Pi$ for $\DBHP_{\cD,\cD'}(n,\beta)$ consists of a message $m$ sent from Alice to Bob. The complexity of the protocol $\Pi$ is the length of the message $m$: $|\Pi|:=|m|$. The goal of the players is to distinguish between the distributions $\cD$ and $\cD'$, and the success probability of $\Pi$ is defined as $\Pr_{(M,w)\sim\cD}[\text{Bob outputs }\cD]/2+\Pr_{(M,w)\sim\cD'}[\text{Bob outputs }\cD']/2$.
\end{definition}

\cite{KKS15} showed that for any constant $\delta>0$, any protocol that solves $\DBHP_{\yes,\ono}(n,\beta)$ with success probability $(1/2+\delta)$ requires $\Omega(\beta^{3/2}\sqrt{n})$ bits of communication. The next lemma shows that the same lower bound extends to the $\DBHP_{\yes,\no}$ problem by an application of the triangle inequality.

\begin{restatable}[A modification of {\cite[Lemma~5.1]{KKS15}}]{lemma}{triangle}
\label{lem:dbhp}
Let $\beta\in(n^{-1/10},1/16)$ and $s\in(n^{-1/10},1)$ be parameters. Any protocol $\Pi$ for $\DBHP_{\yes,\no}(n,\beta)$ that uses $s\sqrt{n}$ bits of communication cannot distinguish between the $\yes$ and $\no$ distributions with success probability more than $1/2+c\cdot(\beta^{3/2}+s)$ for some constant $c>0$ and all large enough $n$.
\end{restatable}

For completeness, we present a proof of Lemma~\ref{lem:dbhp} in Section~\ref{sec:proof of dbhp}. For ease of exposition, now we will use $\DBHP(n,\beta)$ to denote $\DBHP_{\yes,\no}(n,\beta)$.

Finally, note that the graph $G$ in the definition of $\DBHP$ is extremely sparse (in expectation it has $r\approx\beta n< 0.1n$ edges), and, thus, it is not immediately useful for designing hard instances of \mtwocsp{} problems. In order to overcome this issue, \cite{KKS15} used $\DBHP$ where Bob receives a collection of $T$ messages all sampled either from the $\yes$ or $\no$ distribution. Now the union of the $T$ sparse graphs received by Bob can be used in reductions to \mtwocsp{}s.
\begin{definition}[$\DBHP$ with $T$ messages]
For any $\beta\in(0,1/16)$ and $n,T\in\N$, we define \emph{$\DBHP(n,\beta,T)$} as follows. Let $X^*\in\{0,1\}^n$ be a uniformly random vector, and for $1\leq t\leq T$, let $G_i$ be a random graph sampled from $G(n,2\beta/n)$, and $M_i$ be the edge-vertex incidence matrix of~$G_i$. Alice receives $X^*$, and Bob receives a list $(M_1,w_1),\ldots,(M_T,w_T)$, where with probability $1/2$ all $w_t=M_tX^*$ ($\yes$ case), and with probability $1/2$ all $w_t=\mathbf{1}+M_tX^*$ ($\no$ case). The goal of the players is to have a non-trivial advantage over a random guess in distinguishing between the two distributions, while only communication from Alice to Bob is allowed.
\end{definition}

\paragraph{Reduction from $\DBHP$.}
A reduction from $\DBHP(n,\beta,T)$ to $\mcsp(\cG)$ is defined by a pair of algorithms, $\cA$ and $\cB$. Alice receives her input vector $X^*\in\{0,1\}^n$, runs $\cA$ on the input $X^*$, and outputs a set of $\mcsp(\cG)$-clauses. Bob receives a collection of $T$ pairs $(M_t,w_t)$, applies $\cB$ to each of them, and outputs $T$ sets of $\mcsp(\cG)$-clauses. Finally, the resulting instance of the $\mcsp(\cG)$ problem is the union of clauses from $\cA(X^*), \cB(M_1,w_1),\ldots,\cB(M_T,w_T)$.

The reduction above naturally induces two distributions $\cD^Y(\beta,T,\cA,\cB)$ and $\cD^N(\beta,T,\cA,\cB)$ of $\mcsp(\cG)$ instances, corresponding to the $\yes$ and $\no$ distributions of $(M_t,w_t)$. Let us pick some $v^Y$ and $v^N$, such that $\Pr_{\Psi\sim\cD^Y}[\val_\Psi\geq v^Y]>1-o(1)$ and $\Pr_{\Psi\sim\cD^N}[\val_\Psi\leq v^N]>1-o(1)$. Note that for any $\alpha>v^N/v^Y$, an $\alpha$-approximate streaming algorithm for $\mcsp(\cG)$ distinguishes the two distributions $\cD^Y(\beta,T,\cA,\cB)$ and $\cD^N(\beta,T,\cA,\cB)$ with high probability. 
The following theorem states that any streaming algorithm that distinguishes these two distributions, requires space $\Omega(\sqrt{n})$. In particular, any streaming $\alpha$-approximation for $\mcsp(\cG)$ requires space at least $\Omega(\sqrt{n})$.

\begin{restatable}[Reduction from $\DBHP$ with $T$ messages]{theorem}{theoremfoursix}\label{thm:dbhp for streaming}
Let $c>0$ be the constant from Lemma~\ref{lem:dbhp}.
For every $T\in\N$, $0<\beta\leq1/(10cT)^{2/3}$, and reduction $(\cA,\cB)$ from \DBHP{} to $\mcsp(\cG)$, any streaming algorithm that distinguishes $\cD^Y(\beta,T,\cA,\cB)$ and $\cD^N(\beta,T,\cA,\cB)$ with success probability at least $3/4$ requires space at least $\frac{1}{40cT}\cdot\sqrt{n}$.
\end{restatable}

The proof of Theorem~\ref{thm:dbhp for streaming} follows the proofs in~\cite{KKS15} by using the standard hybrid argument as well as the data processing inequality for total variation. We postpone the details of the proof of Theorem~\ref{thm:dbhp for streaming} to Section~\ref{sec:proof of dbhp thm}, and first describe reductions from $\DBHP$ to $\meand$ and $\mor$ in Sections~\ref{sec:reduction and} and~\ref{sec:reduction or}, respectively.

\subsection{From \texorpdfstring{$\DBHP$}{DBHP} to \texorpdfstring{$\meand$}{AND}}\label{sec:reduction and}

Now, we describe the reduction from $\DBHP$ to \meand{}. In order to describe the reduction, it suffices to specify the parameters $\beta$ and $T$, and the algorithms $\cA^\cspeand$ and $\cB^\cspeand$. Recall that we associate a vector $X^*\in\{0,1\}^n$ with the set of its ones: $X\subseteq [n], X=\{i\colon X_i=1\}$ . Also, recall that the input of Bob, $(M,w)$, consists of an edge-vertex incidence matrix ${M\in\{0,1\}^{r\times n}}$ and a vector $w\in\{0,1\}^r$. In particular, every row of $M$ has exactly two ones.

\begin{reduction}{Reduction from $\DBHP$ to $\meand$}
\begin{itemize}
\item Let $c>0$ be the constant from Lemma~\ref{lem:dbhp}. For a given error parameter $\epsilon\in(0,1)$, let $T=(10000/\epsilon^2)^3\cdot(10c)^2$ and $\beta=\frac{1}{(10cT)^{2/3}}$ such that $\beta T=10000/\epsilon^2$.
\item $\cA^\cspeand(X^*)$: Sample $\beta nT/4$ independent pairs $(i,j)\in X^*\times\overline{X^*}$, and for each of them output the clause $(x_i\wedge \neg x_j)$.
\item $\cB^\cspeand(M,w)$: Let $r$ be the number of rows in $M$. For each $1\leq k\leq r$ with $w_k=1$, let the $1$s in the $k^{\text{th}}$ row of $M$ be at the $i^\text{th}$ and $j^\text{th}$ positions, then output two clauses: $(x_i\wedge \neg x_j)$ and $(\neg x_i \wedge x_j)$.
\end{itemize}
\end{reduction}

\begin{restatable}{lemma}{dbhpandval}\label{lem:dbhp and val}
For any $\epsilon\in(0,1)$, let $(\beta,T,\cA^\cspeand,\cB^\cspeand)$ be the parameters described in the above reduction. For a $\meand$ instance $\Psi$, let $m_\Psi$ denote the number of clauses in $\Psi$. Then
\begin{align*}
&\Pr_{\Psi\sim\cD^Y(\beta,T,\cA^\cspeand,\cB^\cspeand)}\left[\val_\Psi<\left(\frac{3}{5}-\epsilon\right)\cdot m_\Psi\right]=o(1)
\intertext{and}
&\Pr_{\Psi\sim\cD^N(\beta,T,\cA^\cspeand,\cB^\cspeand)}\left[\val_\Psi>\left(\frac{4}{15}+\epsilon\right)\cdot m_\Psi\right]=o(1) \, .
\end{align*}
\end{restatable}
We prove Lemma~\ref{lem:dbhp and val} in Section~\ref{sec:gap and}.
An immediate corollary of Theorem~\ref{thm:dbhp for streaming} and Lemma~\ref{lem:dbhp and val} is the desired lower bound for streaming approximation of \meand{}.
\begin{corollary}\label{cor:and lb dbhp}
For any constant $\epsilon\in(0,1)$, any streaming algorithm that $(4/9+\epsilon)$-approximates $\meand$ with success probability at least $3/4$ requires $\Omega(\sqrt{n})$ space.
\end{corollary}

\subsection{From \texorpdfstring{$\DBHP$}{DBHP} to \texorpdfstring{$\mor$}{OR}}\label{sec:reduction or}
Now, we describe the reduction from $\DBHP$ to $\mor$. Again, it suffices to specify the parameters $\beta$ and $T$, and the algorithms $\cA^\cspor$ and $\cB^\cspor$.

\begin{reduction}{Reduction from $\DBHP$ to $\cspor$}
\begin{itemize}
\item Let $c>0$ be the constant from Lemma~\ref{lem:dbhp}. For a given error parameter $\epsilon\in(0,1)$, let  $T=(10000/\epsilon^2)^3\cdot(10c)^2$ and $\beta=\frac{1}{(10cT)^{2/3}}$ such that $\beta T=10000/\epsilon^2$.
\item $\cA^\cspor(X^*)$: 
Sample $\frac{\sqrt{2}-1}{2}\cdot\beta nT$ independent copies of $i\in X^*$, and for each of them output the 1-clause $(x_i)$. Sample another $\frac{\sqrt{2}-1}{2}\cdot\beta nT$ independent copies of $j\in \overline{X^*}$, and for each of them output the 1-clause $(\neg x_j)$.
\item $\cB^\cspor(M,w)$: Let $r$ be the number of rows in $M$. For each $1\leq k\leq r$ with $w_k=1$, let the the $1$s in the $k^{\text{th}}$ row of $M$ be at the $i^\text{th}$ and $j^\text{th}$ positions, then output two clauses: $(x_i \vee x_j)$ and $(\neg x_i \vee \neg x_j)$.
\end{itemize}
\end{reduction}

\begin{restatable}{lemma}{dbhporval}
\label{lem:dbhp or val}
For any $\epsilon\in(0,1)$, let $(\beta,T,\cA^\cspor,\cB^\cspor)$ be the parameters described in the above reduction.
For a $\mor$ instance $\Psi$, let $m_\Psi$ denote the number of clauses in $\Psi$. Then
\begin{align*}
&\Pr_{\Psi\sim\cD^Y(\beta,T,\cA^\cspor,\cB^\cspor)}\left[\val_\Psi=m_\Psi\right]=1
\intertext{and}
&\Pr_{\Psi\sim\cD^N(\beta,T,\cA^\cspor,\cB^\cspor)}\left[\val_\Psi>\left(\frac{\sqrt{2}}{2}+\epsilon\right)\cdot m_\Psi\right]=o(1) \, .
\end{align*}
\end{restatable}
The proof of Lemma~\ref{lem:dbhp or val} is presented in Section~\ref{sec:gap or}.
Now, the desired lower bound for any streaming approximations for \mor{} immediately follows from Theorem~\ref{thm:dbhp for streaming} and Lemma~\ref{lem:dbhp or val}.
\begin{corollary}\label{cor:or lb dbhp}
For any constant $\epsilon\in(0,1)$, any streaming algorithm that $(\sqrt{2}/2+\epsilon)$-approximates $\mor$ with success probability at least $3/4$ requires $\Omega(\sqrt{n})$ space.
\end{corollary}

\subsection{Proof of Lemma~\ref{lem:dbhp}}\label{sec:proof of dbhp}
In this section, we show that the hardness of $\DBHP_{\yes,\ono}(n,\beta)$ proved in~\cite{KKS15} can be easily extended to the hardness of $\DBHP_{\yes,\no}(n,\beta)$.

\triangle*
\begin{proof}
Let us consider a protocol $\Pi$ that uses $s\sqrt{n}$ bits of communication to distinguish between the $\yes$ and $\no$ distributions. For an Alice's input $X^*$, we denote the message that Alice sends to Bob by $\Pi(X^*)$. For each $\cD\in\{\yes,\no,\ono\}$, let $\cP_\cD$ be the distribution of $(M,\Pi(X^*),w)$ where $(X^*,M,w)\sim\cD$.

The equation~(12) in~\cite{KKS15}~\footnote{In~\cite{KKS15}, they use $D^1,D^2$ to denote $\cP_\yes$ and $\cP_{\ono}$. They also used $P$ instead of $\Pi$.} shows that in this case
\[
\|\cP_{\yes}-\cP_{\ono}\|_{tvd}=O(\beta^{3/2}+s) \, .
\]

Observe that when $(X^*,M,w)\sim\ono$, both $(M,\Pi(X^*),w)$ and $(M,\Pi(X^*),\mathbf{1}+w)$ are distributed according to $\cP_{\ono}$. (Indeed, in this case $w\in\{0,1\}^r$ is uniformly random, and independent of the choices of $X^*$ and $M$.) Also, from the definitions of the distributions $\yes$ and $\no$, when $(X^*,M,w)\sim\yes$ (and, thus, $(M,\Pi(X^*),w)\sim\cP_{\yes}$), we have that $(M,\Pi(X^*),\mathbf{1}+w)$ is distributed according to $\cP_\no$.

Further, by the data processing inequality (see Proposition~\ref{lem:tvd properties}), adding the constant vector $\mathbf{1}$ to the variable $w$ in $(M,\Pi(X^*),w)$ does not increase the total variation distance. Thus, we have
\[
\|\cP_{\no}-\cP_{\ono}\|_{tvd}\leq\|\cP_{\yes}-\cP_{\ono}\|_{tvd}=O(\beta^{3/2}+s) \, .
\]
Finally, by the triangle inequality (see Proposition~\ref{lem:tvd properties}),
\[
\|\cP_{\yes}-\cP_{\no}\|_{tvd}\leq\|\cP_{\yes}-\cP_{\ono}\|_{tvd}+\|\cP_{\no}-\cP_{\ono}\|_{tvd}=O(\beta^{3/2}+s) \, .
\]
From the definition of the total variation distance, we have that the success probability of Bob in distinguishing $\yes$ from $\no$ is at most $1/2+O(\beta^{3/2}+s)$, which completes the proof.
\end{proof}

\subsection{Proof of Theorem~\ref{thm:dbhp for streaming}}\label{sec:proof of dbhp thm}
Before presenting the proof of Theorem~\ref{thm:dbhp for streaming}, we will show that a streaming algorithm $\ALG$ for distinguishing the distributions $\cD^Y(\beta,T,\cA,\cB)$ and $\cD^N(\beta,T,\cA,\cB)$ can be turned into a protocol for $\DBHP(n,\beta)$.

\begin{lemma}\label{lem:dbhp hybrid}
Let $n,T,s\in\N$, $\beta\in(n^{-1/10},1/16)$, and let $(\cA,\cB)$ be a reduction from \DBHP{} to $\mcsp(\cG)$. Suppose that a streaming algorithm $\ALG$ distinguishes $\cD^Y(\beta,T,\cA,\cB)$ and $\cD^N(\beta,T,\cA,\cB)$ using space $s$ with probability at least $1/2+\Delta$, then there is a one-way protocol for $\DBHP(n,\beta)$ using at most $s$ bits of communication that succeeds with probability at least $1/2+\Delta/(2T)$.
\end{lemma}
\begin{proof}
First we fix the randomness of the algorithm $\ALG$ so that the resulting deterministic algorithm succeeds with probability at least $1/2+\Delta$ (by the averaging argument). Next, by triangle inequality, $\ALG$ can distinguish either (i) $\cD^Y(\beta,T,\cA,\cB)$ and $\cD^{\overline{N}}(\beta,T,\cA,\cB)$ or (ii) $\cD^N(\beta,T,\cA,\cB)$ and $\cD^{\overline{N}}(\beta,T,\cA,\cB)$ with probability at least $1/2+\Delta/2$. Without loss of generality, let us assume it is case (i) while the case (ii) can be analyzed similarly.

Now, for each $i=0, 1, \dots, T$, let $S_i^Y$ (resp., $S_i^{\overline{N}}$) be the state of $\ALG$ after receiving $\cA(X^*), \cB(M_1,w_1), \ldots, \cB(M_i,w_i)$, where the inputs are sampled from the $\yes$ (resp., $\overline{\no}$) distribution. Note that $\{S_i^Y\}$ and $\{S_i^{\overline{N}}\}$ are random variables, and $\|S_0^Y-S_0^{\overline{N}}\|_{tvd}=0$ while the success probability of $\ALG$ guarantees that $\|S_T^Y-S_T^{\overline{N}}\|_{tvd}\geq\Delta/2$. By the hybrid argument and the triangle inequality for the total variation distance (Proposition~\ref{lem:tvd properties}), there exists $i^*\in[T-1]$ such that
\begin{equation}\label{eq:protocol for BHP using ALG}
\|S_{i^*+1}^Y-S_{i^*+1}^{\overline{N}}\|_{tvd}-\|S_{i^*}^Y-S_{i^*}^{\overline{N}}\|_{tvd}\geq\frac{\Delta}{2T} \, .
\end{equation}

This indicates that the $(i^*+1)^{\text{th}}$ inputs (\emph{i.e.}, $\cB(M_i,w_i)$) are sufficient for distinguishing between the $\yes$ and the $\overline{\no}$ cases with non-trivial probability. Specifically, let $\tilde{S}^Y$ (resp. $\tilde{S}^{\overline{N}}$) be the distribution of the states of $\ALG$, when it starts with a state from $S_{i^*}^Y$ and receives one input $\cB(M,w)$ where $(M,w)$ is sampled from the $\yes$ (resp. $\overline{\no}$) distributions.

\begin{claim}\label{claim:protocol for BHP using ALG}
Let $\tilde{S}^Y$ and $\tilde{S}^{\overline{N}}$ be the random variables defined above, then $\|\tilde{S}^Y-\tilde{S}^{\overline{N}}\|_{tvd}\geq\frac{\Delta}{2T}$.
\end{claim}
\begin{proof}
First, by the triangle inequality for the total variation distance (see Proposition~\ref{lem:tvd properties}), we have
\begin{align*}
\|\tilde{S}^Y-\tilde{S}^{\overline{N}}\|_{tvd}\geq\|\tilde{S}^Y-S_{i^*+1}^{\overline{N}}\|_{tvd}-\|\tilde{S}^{\overline{N}}-S_{i^*+1}^{\overline{N}}\|_{tvd} \,. 
\end{align*}
Note that $\tilde{S}^Y=S_{i^*+1}^Y$ by definition, and $\|\tilde{S}^{\overline{N}}-S_{i^*+1}^{\overline{N}}\|_{tvd}\leq\|S_{i^*}^Y-S_{i^*}^{\overline{N}}\|_{tvd}$ by the data processing inequality. Concretely, we apply item 2 of Proposition~\ref{lem:tvd properties} with $X=S_{i^*}^Y$, $Y=S_{i^*}^{\overline{N}}$, $W=\cB(M_{i^*+1},w_{i^*+1})$ where $(M_{i^*+1},w_{i^*+1})\sim\overline{\no}$, and $f=\ALG$). Note that since $(M_{i^*+1},w_{i^*+1})\sim\overline{\no}$ we have $W$ being independent to both $X$ and $Y$. This is the reason why we need to work on $\yes$ versus $\overline{\no}$ rather than directly using $\yes$ versus $\no$. Finally, this together with~\eqref{eq:protocol for BHP using ALG}, gives the desired bound
\begin{align*}
\|\tilde{S}^Y-\tilde{S}^{\overline{N}}\|_{tvd}
\geq\|S_{i^*+1}^Y-S_{i^*+1}^{\overline{N}}\|_{tvd}-\|S_{i^*}^Y-S_{i^*}^{\overline{N}}\|_{tvd}\geq\frac{\Delta}{2T} \, .
\end{align*}
\end{proof}

Finally, we use $\ALG$ to design a protocol for $\DBHP(n,\beta)$. 
Note that Alice and Bob have the description of the algorithm $\ALG$, and, therefore, know the distributions $\tilde{S}^Y$ and $\tilde{S}^{\overline{N}}$. In particular, they both know the value of $i^*$. Moreover, since Alice and Bob know $i^*$, they know the distributions $\tilde{S}^Y$ and $\tilde{S}^{\overline{N}}$.

\begin{algorithm}[H]
	\caption{A protocol for $\DBHP(n,\beta)$ using $\ALG$}\label{alg:protocol for BHP using ALG}
	\begin{algorithmic}[1]
		\Input Alice receives input $X^*$, and Bob receives inputs $(M,w)$.
		\Goal Distinguish between $MX^*=w$ ($\yes$ case) and $MX^*=\mathbf{1}-w$ ($\overline{\no}$ case).
		\State Alice samples a state $S_A$ of $\ALG$ from the distribution $S_{i^*}^Y$, conditioned on the first input being $\cA(X^*)$. 
		Alice sends $S_A$ to Bob. Since $\ALG$ uses $s$ bits of memory,~$|S_A|\leq s$.
		\State Bob executes $\ALG$ with the initial state $S_A$ on the input $\cB(M,w)$. Let $S$ be the resulting state of $\ALG$.
		\State Bob outputs $\yes$ if $\Pr[\tilde{S}^Y=S]\geq\Pr[\tilde{S}^{\overline{N}}=S]$; otherwise Bob outputs $\overline{\no}$.
	\end{algorithmic}
\end{algorithm}
Let $\Omega_Y=\{S:\Pr[\tilde{S}^Y=S]\geq\Pr[\tilde{S}^{\overline{N}}=S]\}$ and $\Omega_{\overline{N}}=\{S:\Pr[\tilde{S}^Y=S]<\Pr[\tilde{S}^{\overline{N}}=S]\}$, then
\begin{align*}
\|\tilde{S}^Y-\tilde{S}^{\overline{N}}\|_{tvd}&=\frac{1}{2}\sum_{S\in\Omega_Y}\Pr[\tilde{S}^Y=S]-\Pr[\tilde{S}^{\overline{N}}=S]+\frac{1}{2}\sum_{S\in\Omega_{\overline{N}}}\Pr[\tilde{S}^{\overline{N}}=S]-\Pr[\tilde{S}^Y=S] \\
&=\left(\sum_{S\in\Omega_Y}\Pr[\tilde{S}^Y=S]-\Pr[\tilde{S}^{\overline{N}}=S]\right)-\frac{1}{2} \, .
\end{align*}
This implies that Bob correctly identifies the $\yes$ distribution with probability
\[
\Pr_{(M,w)\sim\yes}\left[\text{Bob outputs $\yes$}\right]=\sum_{S\in\Omega_Y}\Pr[\tilde{S}^Y=S]\geq\frac{1}{2}+\|\tilde{S}^Y-\tilde{S}^{\overline{N}}\|_{tvd} \, .
\]
Similarly,
\[
\Pr_{(M,w)\sim\overline{\no}}\left[\text{Bob outputs $\overline{\no}$}\right]\geq\frac{1}{2}+\|\tilde{S}^Y-\tilde{S}^{\overline{N}}\|_{tvd} \, .
\]
Thus, the above protocol solves $\DBHP(n,\beta)$ with success probability at least
\begin{align*}
\frac{1}{2}\Pr_{(M,w)\sim\yes}\left[\text{Bob outputs $\yes$}\right]+\frac{1}{2}\Pr_{(M,w)\sim\overline{\no}}\left[\text{Bob outputs $\overline{\no}$}\right]
\geq \frac{1}{2}+\|\tilde{S}^Y-\tilde{S}^{\overline{N}}\|_{tvd}\geq\frac{1}{2}+\frac{\Delta}{2T} \, ,
\end{align*}
where the last inequality is due to Claim~\ref{claim:protocol for BHP using ALG}.
\end{proof}

Now we are ready to finish the proof of Theorem~\ref{thm:dbhp for streaming}.
\theoremfoursix*
\begin{proof}
Consider a streaming algorithm that distinguishes between the distributions  $\cD^Y(\beta,T,\cA,\cB)$ and $\cD^N(\beta,T,\cA,\cB)$ with probability at least $3/4$ using space $S$. 
Then, by Lemma~\ref{lem:dbhp hybrid}, there exists a protocol for $\DBHP(n,\beta)$ with at most $S$ bits of communication, and success probability $1/2+1/(8T)$.

On the other hand, by Lemma~\ref{lem:dbhp}, any protocol for $\DBHP(n,\beta)$ with 
success probability 
\[
\frac{1}{2}+c\cdot\left(\beta^{3/2}+\frac{1}{40cT}\right)\stackrel{\beta\leq1/(10cT)^{2/3}}{\leq}\frac{1}{2}+\frac{1}{8T}
\]
must use  $S\geq \frac{1}{40cT}\cdot\sqrt{n}$ bits of communication.
\end{proof}

\section{Analysis for the gap of \meand~and \mor~instances}\label{sec:gap instances}

The goal of this section is to prove Lemma~\ref{lem:dbhp and val} and Lemma~\ref{lem:dbhp or val}. 
We analyse the structure of $\DBHP$ in Section~\ref{sec:graphical view} and present an intuitive and graphical view of the reductions. After that, we give the proofs for Lemma~\ref{lem:dbhp and val} and Lemma~\ref{lem:dbhp or val} in Section~\ref{sec:gap and} and Section~\ref{sec:gap or} respectively.

\paragraph{Notation for an assignment.}
In this section, we interchangeably work with one of the following representations for $\sigma$ in order to simplify the presentation. Previously, $\sigma$ was defined as a function that maps $\{x_1,x_2,\dots,x_n\}$ to $\{0,1\}$. It can be represented by a vector in $\{0,1\}^n$ which has $\sigma(x_i)$ as its $i^\text{th}$ coordinate. It can also be represented by the set $\{i\in[n]: \sigma(x_i)=1 \}$.

\subsection{A graphical view of \DBHP}\label{sec:graphical view}

Here we introduce a graphical view of $\DBHP$ which will provide a more intuitive lens to understand the reductions. Recall that in $\DBHP$, Bob has private inputs $M\in\{0,1\}^{r}$ and $w\in\{0,1\}^{r}$, where $M$ is the edge-incidence matrix of an $n$-vertex graph $G$ and $w$ is an indicator vector. Specifically, $M$ corresponds to a graph sampled from $G(n,2\beta/n)$ and $r$ denotes the number of edges in this graph. We focus on the subgraph $H\subseteq G$ that contains only those edges from $M$ whose corresponding entry in $w$ is $1$. We examine the distributions of this subgraph $H$ under different input distributions to $\DBHP$. Recall that we are interested in two input distributions to $\DBHP$: $\yes$ and $\no$. In both of these distributions, we first sample a hidden partition $X^*\in\{0,1\}^*$ and then sample $T$ independent graphs from $G(n,2\beta/n)$ where the edge-vertex incidence matrices of these graphs are denoted as $\{M_t\}_{t\in[T]}$. In the $\yes$ distribution, $w_t = M_t X^*$ and in the $\no$ distribution, $w_t = \mathbf{1} - M_tX^*$. We will abuse notation and call the corresponding distributions of the subgraph $H$ as $\yes$ and $\no$ respectively. We summarize the properties of these distributions in the following lemma.

\begin{figure}[ht]
    \centering
    \includegraphics[width=8cm]{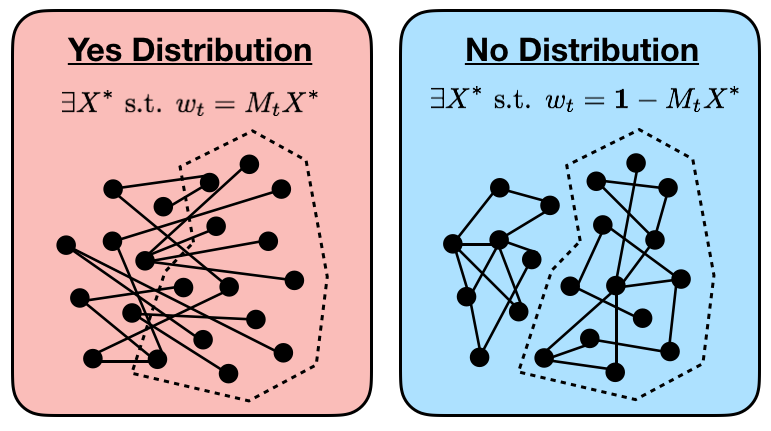}
    \caption{For a random graph on vertex set $[n]$, we partition the edges into two sets: (i) edges that lie across $X^*$ and $\overline{X^*}$ and (ii) edges that lie in $X^*$ or $\overline{X^*}$. In the $\yes$ distribution, only the (i) type edges are present in~$H$. In the $\no$ distribution, only the (ii) type edges are present in~$H$. }
    \label{fig:graphical view}
\end{figure}

\begin{lemma}[Graphical view of $\DBHP$]\label{lem:graphical view}
For any $n\in\N$ large enough and $\epsilon\in(0,0.25)$, let ${T=(10000/\epsilon^2)^3\cdot(10c)^2}$ and $\beta=\frac{1}{(10cT)^{2/3}}$ such that $\beta T=10000/\epsilon^2$. Let $\yes$ and $\no$ be the distributions of the subgraph $H$ induced from $\DBHP(n,\beta,T)$ as described above, and let $m_\DBHP$ denote the total number of edges in~$H$. For every $X^*,\sigma\in\{0,1\}^n$, define $m_\text{cross}(\sigma)$ to be the number of edges $(i,j)$ such that (i) $\sigma(x_i)\neq\sigma(x_j)$ and (ii) $X^*_i=X^*_j$. We have the following.
\begin{align*}
\intertext{$\bullet$ (Size of $X^*$) For each distribution $\yes,\no$ and for any constant $\epsilon'\in(0,1)$ such that $\epsilon'\geq\epsilon/10$, we have}
&\Pr\left[\left||X^*|-\frac{n}{2}\right|>\epsilon'\cdot n\right]=o(1) \, .
\intertext{$\bullet$ (Number of edges) For each distribution $\yes,\no$ and for any constant $\epsilon'\in(0,1)$ such that $\epsilon'\geq\epsilon/10$, we have}
&\Pr\left[\left|m_\DBHP-\frac{\beta nT}{2}\right|>\epsilon'\cdot\beta nT\right]=o(1) \, .
\intertext{$\bullet$ ($\no$ distribution) For any constant $\epsilon'\in(0,1)$ such that $\epsilon'\geq\epsilon/10$, we have}
&\Pr_{\no}\left[\exists\sigma\in\{0,1\},\ m_\text{cross}(\sigma)>\left(\frac{|\sigma\cap X^*|\cdot|\overline{\sigma}\cap X^*|+|\sigma\cap\overline{X^*}|\cdot|\overline{\sigma}\cap\overline{X^*}|}{n^2}\right)\cdot2\beta nT+\epsilon'\cdot \beta nT\right]=o(1) \, .
\end{align*}
\end{lemma}
\begin{proof}
\mbox{}
\begin{itemize}
\item (Size of $X^*$) For any $\epsilon'\in(0,1)$, we have
\[
\Pr\left[\left||X^*|-\frac{n}{2}\right|>\epsilon'\cdot n\right]\leq\frac{2\cdot\sum_{i=0}^{\ceil*{n/2-\epsilon'\cdot n}}\binom{n}{i}}{2^n}=2^{-\Omega_{\epsilon'}(n)}=o(1) \, .
\]

\item (Number of edges) Here we only prove the case for $\yes$ distribution while the other case can be proved similarly. Also, we only show the upper bound for $m_\DBHP-\beta nT/2$ while the lower bound can be proved symmetrically.

Let $\epsilon''\in(0,1)$ be error parameters that will be chosen in the end according to $\epsilon'$. First, by Lemma~\ref{lem:chernoff}, we have for any constant $\epsilon''\in(0,1)$,
\begin{equation}\label{eq:graphical view X star}
\Pr_{\yes}\left[\left||X^*|-\frac{n}{2}\right|>\epsilon''\cdot n\right]=o(1) \, .
\end{equation}
Denote the event where $-\epsilon''n\leq|X^*|-n/2\leq\epsilon''n$ as $\GOOD$.
Now, for each $t\in[T]$, let $m_t=\|M_tX^*\|_1$, \textit{i.e.,} the number of edges in $M_t$ that crosses $X^*$ and $\overline{X^*}$, we have
\[
\Exp_\yes\left[m_t\ |\ \GOOD\right]\leq \left(\frac{1}{2}+\epsilon''\right)\cdot\beta n \, .
\]
Further, note that when conditioning on $X^*$, $m_t$ are independent, thus by Chernoff bound, when $n$ is large enough, we have
\[
\Pr_{\yes}\left[\sum_{t\in[T]}m_t-\left(\frac{1}{2}+\epsilon''\right)\cdot\beta nT>\epsilon''\cdot\frac{\beta nT}{2}\ \Big|\ \GOOD\right]=o(1) \, .
\]
As $m_\DBHP=\sum_{t\in[T]}m_t$, by choosing $\epsilon'' = \epsilon'/3$, we conclude that
\[
\Pr\left[\left|m_\DBHP-\frac{\beta nT}{2}\right|>\epsilon'\cdot\frac{\beta nT}{2}\right]=o(1) \, .
\]

\item ($\no$ distribution) First, for each $t\in[T]$, let $m^{(t)}_\text{cross}(\sigma)$ denote the number of cross edges (\textit{i.e., $\sigma(x_i)\neq\sigma(x_j)$ and $X_i^*=X_j^*$}) from $M_t.$ Note that $m_\text{cross}=\sum_{t\in[T]}m^{(t)}_\text{cross}(\sigma)$. Next, observe that for each $t\in[T]$, $m^{(t)}_\text{cross}(\sigma)$ is a sum of $|\sigma\cap X^*|\cdot|\overline{\sigma}\cap X^*|+|\sigma\cap\overline{X^*}|\cdot|\overline{\sigma}\cap\overline{X^*}|$ independent Bernoulli random variables with expectation $2\beta/n$. Also, random variables $m^{(1)}_\text{cross}(\sigma),\dots,m^{(T)}_\text{cross}(\sigma)$ are independent to each other because we have fixed $X^*$ and $\sigma$.
Thus, by Chernoff bound (\textit{i.e.,}~Lemma~\ref{lem:chernoff}), we have
\begin{align*}
&\Pr\left[m_\text{cross}(\sigma)>\left(|\sigma\cap X^*|\cdot|\overline{\sigma}\cap X^*|+|\sigma\cap\overline{X^*}|\cdot|\overline{\sigma}\cap\overline{X^*}|\right)\cdot\frac{2\beta}{n} \cdot T+\epsilon'\beta nT\right]\\
<\ &\exp\left(-\frac{\epsilon'^2\beta^2n^2T^2}{2\beta nT}\right)=\exp\left(-\frac{\epsilon'^2\beta nT}{2}\right)<2^{-10 n}
\end{align*}
where the last inequality is due to the choice of $T=10000/\epsilon^2$ and $\epsilon'\geq\epsilon/10$.
Finally, by union bound, we have
\[
\Pr\left[\exists\sigma\in\{0,1\}^n,\ m_\text{cross}(\sigma)>\left(\frac{|\sigma\cap X^*|\cdot|\overline{\sigma}\cap X^*|+|\sigma\cap\overline{X^*}|\cdot|\overline{\sigma}\cap\overline{X^*}|}{n^2}\right)\cdot2\beta nT+\epsilon'\beta nT\right]=o(1) \, .
\]

\end{itemize}
\end{proof}

\subsection{The gap of \meand~instances}\label{sec:gap and}
In this subsection, we complete the proof of the following lemma using the graphical view of $\DBHP$.
\dbhpandval*
\begin{proof}

Recall that $\cA^\cspeand(X^*)$ uses $X^*$ to sample $\beta nT/4$ independent copies of $(i,j)\in X^*\times\overline{X^*}$ and outputs $(x_i\wedge \neg x_j)$. On the other hand, for each row of $M$ with the $i^\text{th}$ and $j^\text{th}$ entry being $1$, if the corresponding entry of $w$ is $1$, $\cB^\cspeand(M,w)$ outputs $(x_i\wedge \neg x_j)$ and $(\neg x_i \wedge x_j)$.

\begin{itemize}
\item (\yes~distribution) Consider the following assignment $\sigma$: \[\sigma(x_i) = \begin{cases}
1, i\in X^* \, ;\\
0, \text{otherwise}\, .
\end{cases} \] Under this assignment, each clause $(x_i\wedge \neg x_j)$ generated by $\cA^\cspeand$ is satisfied since $(i,j)\in X^*\times \overline{X^*}$. For every pair of clauses $(x_i\wedge \neg x_j)$ and $(\neg x_i \wedge x_j)$ generated by $\cB^\cspeand$, exactly one of them is satisfied by $\sigma$. Therefore, $\sigma$ satisfies $\beta nT/4+m_\DBHP$ clauses while the total number of clauses is $m_\Psi=\beta nT/4+2m_\DBHP$.

From Lemma~\ref{lem:graphical view}, we know that $m_\DBHP\in(1\pm\epsilon/10)\cdot\beta nT/2$ with probability at least $1-o(1)$. Thus, we have $m_\Psi\in(5/4\pm\epsilon/10)\cdot\beta nT$ and $\sigma$ satisfies at least $(1-\epsilon/10)\cdot\beta nT$ clauses with probability $1-o(1)$. Therefore, $\val(\sigma)\geq(3/5-\epsilon)\cdot m_\Psi$ with probability at least $1-o(1)$.

\item (\no~distribution) Consider any fixed assignment $\sigma\in\{0,1\}^n$: $\sigma$ satisfies a clauses $(x_i\wedge \neg x_j)$ generated by $\cA^\cspeand$ if and only if $\sigma(x_i)=1$ and $\sigma(x_j)=0$. Let $a(\sigma)$ be the random variable that denotes the number of clauses generated by $\cA^\cspeand$ which are satisfied by $\sigma$. Observe that $a(\sigma)$ is the sum of $\beta nT/4$ independent Bernoulli random variables with mean $|\sigma\cap X^*|\cdot|\overline{\sigma}\cap\overline{X^*}|/(|X^*|\cdot|\overline{X^*}|)$.By Chernoff bound (\textit{i.e.,}~Lemma~\ref{lem:chernoff}), we have
\[
\Pr_{\no}\left[a(\sigma)>\frac{|\sigma\cap X^*|\cdot|\overline{\sigma}\cap\overline{X^*}|}{|X^*|\cdot|\overline{X^*}|}\cdot\frac{\beta nT}{4}+\frac{\epsilon\beta nT}{10}\right]<2^{-10n} \, .
\]
Applying the union bound, we have
\begin{equation}\label{eq:and lb alice}
\Pr_{\no}\left[\exists\sigma\in\{0,1\}^n,\ a(\sigma)>\frac{|\sigma\cap X^*|\cdot|\overline{\sigma}\cap\overline{X^*}|}{|X^*|\cdot|\overline{X^*}|}\cdot\frac{\beta nT}{4}+\frac{\epsilon\beta nT}{10}\right]=o(1) \, .
\end{equation}

Now, let us consider the clauses generated by $\cB^\cspeand$: an edge $(i,j)$ in $M_t$ is selected by $w_t$ if and only if $X^*$ contains both $i$ and $j$, or contains neither (\textit{i.e.}, $X^*_i=X^*_j$). Observe that exactly one of $(x_i\wedge \neg x_j)$ and $(\neg x_i\wedge x_j)$ is satisfied by $\sigma$ if and only if $\sigma(x_i)\neq\sigma(x_j)$; otherwise, both are unsatisfied. Therefore, the number of clauses satisfied by $\sigma$ is exactly $m_\text{cross}(\sigma)$, \textit{i.e.,} the number of edges $(i,j)$ such that (i) $X^*_i=X^*_j$ and (ii) $\sigma(x_i)\neq\sigma(x_j)$. Therefore, the total number of satisfied clauses is given by
\begin{equation}\label{eq:and lb value}
\val_\Psi(\sigma)=a(\sigma)+m_\text{cross}(\sigma) \, .
\end{equation}

By Lemma~\ref{lem:graphical view}, we have
\begin{equation*}
\Pr_\no\left[\exists\sigma\in\{0,1\}^n,\ m_\text{cross}(\sigma)>\left(\frac{|\sigma\cap X^*|\cdot|\overline{\sigma}\cap X^*|+|\sigma\cap\overline{X^*}|\cdot|\overline{\sigma}\cap\overline{X^*}|}{n^2}\right)\cdot2\beta nT+\frac{\epsilon\beta nT}{10}\right]=o(1) \, .
\end{equation*}
Since $|X^*|\cdot |\overline{X^*}| \leq \frac{n^2}{4}$,
\begin{equation}\label{eq:and lb bob}
\Pr_\no\left[\exists\sigma\in\{0,1\}^n,\ m_\text{cross}(\sigma)> \left(\frac{|\sigma\cap X^*|\cdot|\overline{\sigma}\cap X^*|+|\sigma\cap\overline{X^*}|\cdot|\overline{\sigma}\cap\overline{X^*}|}{|X^*|\cdot |\overline{X^*}|}\right)\cdot\frac{\beta nT}{2}+\frac{\epsilon\beta nT}{10}\right]=o(1) \,
\end{equation}

Let $p=|\sigma\cap X^*|/|X^*|\in[0,1]$ and $q=|\overline{\sigma}\cap\overline{X^*}|/|\overline{X^*}|\in[0,1]$.

Combining~\eqref{eq:and lb alice},~\eqref{eq:and lb value}, and~\eqref{eq:and lb bob}, we get
\[
\Pr_{\Psi\sim\cD^N(\beta,T,\cA^\cspeand),\cB^\cspeand}\left[\exists\sigma\{0,1\}^n,\ \val_\Psi(\sigma)>\left(\frac{pq+2p(1-p)+2q(1-q)}{4}\right)\cdot\beta nT+\frac{\epsilon\beta nT}{5}\right]=o(1) \, .
\]

We have $\frac{pq+2p(1-p)+2q(1-q)}{4}=\frac{8-(3p-2)^2-(3q-2)^2-3(p-q)^2}{24}\leq1/3$. Since $m_\Psi\in(5/4\pm\epsilon/10)\cdot\beta nT$ with probability $1-o(1)$, we conclude that
\[
\Pr_{\Psi\sim\cD^N(\beta,T,\cA^\cspeand),\cB^\cspeand}\left[\exists\sigma\{0,1\}^n,\ \val_\Psi(\sigma)>\left(\frac{4}{15}+\epsilon\right)\cdot m_\Psi\right]=o(1) \, .
\]
\end{itemize}
\end{proof}

\subsection{The gap of \mor~instances}\label{sec:gap or}
In this subsection, we complete the proof of the following lemma using the graphical view of $\DBHP$.
\dbhporval*
\begin{proof}
Recall that $\cA^\cspor(X^*)$ uses $X^*$ to sample $\frac{\sqrt{2}-1}{2}\cdot\beta nT$ independent copies of $i\in X^*$ and another $\frac{\sqrt{2}-1}{2}\cdot\beta nT$ independent copies of $j\in \overline{X^*}$ and output $(x_i)$ as well as $(\neg x_j)$. On the other hand, for each row of $M$ with the $i^\text{th}$ and $j^\text{th}$ entry being $1$, if the corresponding entry of $w$ is $1$, $\cB^\cspor$ outputs $(x_i\vee x_j)$ and $(\neg x_i \vee \neg x_j)$.

\begin{itemize}
\item (\yes~distribution) Consider the following assignment $\sigma$: \[\sigma(x_i) = \begin{cases}
1, i\in X^* \, ;\\
0, \text{otherwise} \, .
\end{cases}\]. Under this assignment, every clause of the form $(x_i)$ or of the form $(\neg x_j)$ generated by $\cA^\cspor$ is satisfied because $i\in X^*$ and $j\in\overline{X^*}$. Similarly, every pair of clauses $(x_i\vee x_j)$ and $(\neg x_i\vee\neg x_j)$ generated by $\cB^\cspor$ are also satisfied since in the \yes~distribution, $(i,j)\in X^* \times \overline{X^*}$. Thus, $\val_\Psi=m_\Psi$ as desired.

\item (\no~distribution) Consider any fixed assignment $\sigma\in\{0,1\}^n$: Let $a(\sigma)$ be the random variable that denotes the number of clauses generated by $\cA^\cspeand$ which are satisfied by $\sigma$. Observe that $a(\sigma)$ is the sum of $\frac{\sqrt{2}-1}{2}\beta nT\geq100n$ independent Bernoulli random variables with mean $|\sigma\cap X^*|/|X^*|$ and $\frac{\sqrt{2}-1}{2}\beta nT\geq100n$ independent Bernoulli random variables with mean $|\overline{\sigma}\cap \overline{X^*}|/|\overline{X^*}|$. By Chernoff bound (\textit{i.e.,}~Lemma \ref{lem:chernoff}), we have
\[
\Pr_{\no}\left[a(\sigma)>\left(\frac{|\sigma\cap X^*|}{|X^*|}+\frac{|\overline{\sigma}\cap\overline{X^*}|}{|\overline{X^*}|}\right)\cdot \frac{\sqrt{2}-1}{2}\beta nT+\frac{\epsilon\beta nT}{15}\right]<2^{-10n} \, .
\]

Applying the union bound, we have
\begin{equation}\label{eq:or lb alice}
\Pr_\no\left[\exists\sigma\in\{0,1\}^n,\ a(\sigma)>\left(\frac{|\sigma\cap X^*|}{|X^*|}+\frac{|\overline{\sigma}\cap\overline{X^*}|}{|\overline{X^*}|}\right)\cdot \frac{\sqrt{2}-1}{2}\beta nT+\frac{\epsilon\beta nT}{15}\right]=o(1) \, .
\end{equation}

Now, consider the clauses generated by $\cB^\cspor$: an edge $(i,j)$ in $M_t$ is selected by $w_t$ if and only if $X_i^*=X_j^*$. Observe that both the clauses $(x_i\vee x_j)$ and $(\neg x_i \vee \neg x_j)$ are satisfied if and only if $\sigma(x_i)\neq\sigma(x_j)$; otherwise, exactly one of them is satisfied. Therefore, the number of satisfied clauses among the clauses generated by $\cB^\cspor$ is $m_\DBHP+m_\text{cross}(\sigma)$. Therefore, the total number of satisfied clauses is given by
\begin{equation}\label{eq:or lb value}
\val_\Psi(\sigma) = a(\sigma)+m_\DBHP+m_\text{cross}(\sigma) \, .
\end{equation}

By~Lemma~\ref{lem:graphical view}, we have
\begin{equation*}
\Pr_\no\left[\exists\sigma\in\{0,1\}^n,\ m_\text{cross}(\sigma)>\left(\frac{|\sigma\cap X^*|\cdot|\overline{\sigma}\cap X^*|+|\sigma\cap\overline{X^*}|\cdot|\overline{\sigma}\cap\overline{X^*}|}{n^2}\right)\cdot2\beta nT+\frac{\epsilon\beta nT}{15}\right]=o(1) \, .
\end{equation*}
and 
\begin{equation}\label{eq:or m_DBHP}
\Pr_\no\left[ m_\DBHP > \frac{\beta nT}{2} + \epsilon\cdot\frac{\beta nT}{15}\right]=o(1).    
\end{equation}

Since $|X^*|\cdot |\overline{X^*}| \leq \frac{n^2}{4}$,
\begin{equation}\label{eq:or lb bob}
\Pr_\no\left[\exists\sigma\in\{0,1\}^n,\ m_\text{cross}(\sigma)> \left(\frac{|\sigma\cap X^*|\cdot|\overline{\sigma}\cap X^*|+|\sigma\cap\overline{X^*}|\cdot|\overline{\sigma}\cap\overline{X^*}|}{|X^*|\cdot |\overline{X^*}|}\right)\cdot\frac{\beta nT}{2}+\frac{\epsilon\beta nT}{15}\right]=o(1) \,
\end{equation}

Let $p=|\sigma\cap X^*|/|X^*|\in[0,1]$ and $q=|\overline{\sigma}\cap\overline{X^*}|/|\overline{X^*}|\in[0,1]$. Combining ~\eqref{eq:or lb alice},~\eqref{eq:or lb value},~\eqref{eq:or lb bob} and~\eqref{eq:or m_DBHP}, we get
\[
\Pr_{\Psi\sim\cD^N(\beta,T,\cA^\cspor),\cB^\cspor}\left[\exists\sigma\{0,1\}^n,\ \val_\Psi(\sigma)>\left(\frac{(p+q)(\sqrt{2}-1)}{2}+\frac{1}{2}+\frac{p(1-p)+q(1-q)}{2}\right)\cdot\beta nT+\frac{\epsilon\beta nT}{5}\right]=o(1) \, .
\]

We have $\frac{p+q}{2}\cdot\left(\sqrt{2}-1\right)+\frac{1}{2}+\frac{p(1-p)+q(1-q)}{2}=1-\frac{(p-\sqrt{2}/2)^2+(q-\sqrt{2}/2)^2}{2}\leq1$. Since $m_\Psi\in(\sqrt{2}\pm\epsilon/10)\cdot\beta nT$ with probability $1-o(1)$, we conclude that
\[
\Pr_{\Psi\sim\cD^N(\beta,T,\cA^\cspor),\cB^\cspor}\left[\exists\sigma\{0,1\}^n,\ \val_\Psi(\sigma)>\left(\frac{\sqrt{2}}{2}+\epsilon\right)\cdot m_\Psi\right]=o(1) \, .
\]

\end{itemize}

\end{proof}

%% file: theorem.tex
\section{Proof of Theorem~\ref{thm:main}}
\label{sec:theorem}

\thmmain*

\begin{table}[ht]
\centering
\resultstable
\renewcommand\thetable{1}
\caption{Summary of known and new approximation factors $\alpha_\cG$ for  $\mcsp(\cG)$. We have suppressed~$(1\pm\epsilon)$ multiplicative factors.}
\end{table}

\begin{proof}
Note that for $\cG$ listed in Table~\ref{table:results}, the space lower bounds for $\mcsp{\cG}$ are proven in Corollary~\ref{cor:eor lb}, Corollary~\ref{cor:or lb dbhp}, Theorem~\ref{thm:kk19}, and Corollary~\ref{cor:and lb dbhp}, respectively. Then the space lower bound for any $\mcsp(\cF)$ directly follows from the fact that for $\cG \subseteq \cF$, any hard instance for $\mcsp(\cG)$ is also a hard instance of $\mcsp(\cF)$. 

We provide a case-by-case analysis to prove the upper bounds.

\paragraph*{Case I -- $\arg \min_{\cG \subseteq \cF} \alpha_{\cG} \in \left\{\cspt , \cspor , \{\cspt, \cspor\}\right\}$:}
In this case, $\cF= \{\cspor\}$, $\cF= \{\cspt\}$, or $\cF= \{\cspor, \cspt\}$, and each of these cases is covered in Table~\ref{table:results}. The corresponding upper bounds for these cases are proven in Proposition~\ref{thm:trivial alg} and Theorem~\ref{thm:alg or}.

\paragraph*{Case II -- $\arg \min_{\cG \subseteq \cF} \alpha_{\cG} = \cspxor$:}
In this case, $\cF\subseteq \{\cspor,\cspxor,\cspt\}$. Consider any instance $\Psi$ of $\mcsp(\cF)$. A random assignment satisfies every constraint in $\Psi$ with probability at least $1/2$. Therefore, the trivial streaming algorithm that counts the number of clauses, $m$ and outputs $m/2$ achieves $1/2$ approximation (see Proposition~\ref{thm:trivial alg}).

\paragraph*{Case III -- $\arg \min_{\cG \subseteq \cF} \alpha_{\cG} = \cspand$:}
In this case, $\cF\subseteq \{\cspand,\cspor,\cspxor,\cspt\}$. Any Boolean constraint $f(x)$ of length at most $2$ can be expressed as the disjunction of (at most $4$) $\cspand$ constraints $\{f_i(x)\}$ such that for any assignment $\sigma$, the number of satisfied constraints among $\{f_i(\sigma)\}$ is exactly $1$ if $f(\sigma)=1$; otherwise, it is $0$. 
Therefore, any instance $\Psi$ of $\mcsp(\cF)$ can be reduced to $\Psi'$, an instance of \mand{} with the same optimal value. Now the approximation algorithm for \mand{} from Theorem~\ref{thm:alg and} finishes the proof.
\end{proof}

%% file: maxksat.tex
\section{A tight approximation algorithm for \texorpdfstring{$\mksat$}{Max-kSAT} }\label{sec:mksat}

It follows from Theorem \ref{thm:main} that for every $\epsilon >0$, any $(\sqrt{2}/2+\epsilon)$-approximate streaming algorithm for $\mksat$ requires space $\Omega(\sqrt{n})$ since any $\mksat$ instance is also an instance of $\msat{}$. In this section, we prove that this lower bound is indeed tight, \textit{i.e.,} we show that for any $\epsilon>0$, there exists a streaming algorithm that uses space $\mathcal{O}(\epsilon^{-2}\log n)$ and computes $(\sqrt{2}/2-\epsilon)$-approximation for $\mksat$.
\\

We first extend the notion of bias which we defined in Section \ref{sec:prelims} to $\mksat$ instances. Let $\Psi$ be an instance of $\mksat$ with $m=|\Psi|$ clauses. For $r\in \mathbb{N}$, an $r$-clause is a clause that depends on $r$ variables. We denote the total number of $r$-clauses by $m_r$, and the total number of clauses that depend on \textit{at least} $r$ variables by $m_{\ge r}$. We use $\posvar_i^{(r)}(\Psi)$ for the number of $r$-clauses where the variable $x_i$ appears positively. Similarly, $\negvar_i^{(r)}(\Psi)$ denotes the number of $r$-clauses containing $\neg x_i$.

\begin{definition}
The \emph{bias} of a variable $x_i$ of an instance $\Psi$ of $\mksat$ is defined as 
\[
\bias_i(\Psi) = \left|\sum_{r} \frac{1}{2^{r-1}} \left(\posvar_i^{(r)}(\Psi) - \negvar_i^{(r)}(\Psi)\right)\right|
\, .
\]
The {bias vector} of $\Psi$ is a vector $\bm{b}\in\Real^n$, where $\bm{b}_i=\bias_i(\Psi)$.
Finally, the \emph{bias} of the formula $\Psi$ is defined as the sum of biases of its variables:
\[
 \bias(\Psi) = \sum_{i=1}^n \bias_i(\Psi) = \sum_{i=1}^n \left|\sum_{r} \frac{1}{2^{r-1}} \left(\posvar_i^{(r)}(\Psi) - \negvar_i^{(r)}(\Psi)\right)\right| \, .
\]
\end{definition}

In Lemmas~\ref{lem:ubSAT} and~\ref{lem:ksat bias lower bound} we give upper and lower bounds on $\val_\Psi$ in terms of $\bias(\Psi)$ and $m_r$. We postpone their proofs to Sections~\ref{sec:proofs_ksat ub} and~\ref{sec:proofs_ksat lb}. In this section we prove that the ratio between the presented lower and upper bounds is bounded by $\frac{\sqrt{2}}{2}$, and that there is a $O(\log{n})$-space algorithm that sketches the lower bounds of Lemma~\ref{lem:ksat bias lower bound}  on $\val_\Psi$. The following lemma gives an upper bound on $\val_\Psi$.

\begin{restatable}{lemma}{ubk}
\label{lem:ubSAT}
Let $\Psi$ be a $\mksat$ instance. Then
\[
\val_{\Psi} \leq \min \left\{\sum_{j=1}^k m_j, \frac{\bias(\Psi)}{2} + \sum_{j=1}^k  \left(\frac{2^j+j-2}{2^j}\right)m_j \right\} 
\, .
\]
\end{restatable}

For lower bounds, the trivial algorithm guarantees that for every $\mksat$ instance $\Psi$, $\val_\Psi\geq \sum_{j=1}^k (1-2^{-j}) m_j$. This bound is not sufficient for proving tight space lower bounds for streaming algorithms, so we improve it as follows. For instances with high bias, we prove a (stronger) lower bound of $\val \geq \frac{\bias(\Psi)}{2} + \sum_{j=1}^k \frac{j}{2^j} m_j $. In order to handle the case of low bias, we design a distribution of assignments which in expectation satisfy a large number of clauses in formulas with low $\bias$. To summarize, we have the following lemma on the lower bound for $\val_\Psi$.

\begin{restatable}{lemma}{biaslbk}
\label{lem:ksat bias lower bound}
Let $\Psi$ be a $\mksat$ instance. Then
\begin{enumerate}[leftmargin=*]
    \item \(\val_\Psi \geq \frac{\bias(\Psi)}{2} + \sum_{j=1}^k \frac{j}{2^j} m_j \, ;\)
    \item if $\bias(\Psi)\leq \sum_{j=2}^k \frac{2^j-j-1}{2^{j-2}} m_j$, then
    \[\val_\Psi \geq \sum_{j=1}^k (1-2^{-j})m_j +\frac{\bias(\Psi)^2}{4\left(\sum_{j=2}^k (\frac{2^j-j-1}{2^{j-2}})m_j\right)}  \, .\]
\end{enumerate}
\end{restatable}

Now we are ready to present an approximation algorithm for the $\mksat$ problem.

\begin{algorithm}[ht] 
	\caption{$(\sqrt{2}/2-\epsilon)$-approximation streaming algorithm for $\mksat)$}\label{alg:mSAT}
	\begin{algorithmic}[1]
	    \Input $\Psi$---an instance of $\mksat$. Error parameter $\epsilon \in (0,0.01)$.
		\State Approximate the $\ell_1$-norm of the bias vector with error $\delta= \epsilon/8$ (Theorem~\ref{thm:l1sketch}):
		\Statex Compute $B\in\left(1\pm \delta\right)\bias(\Psi)$.
		\State Use $k$ counters to count the number of $j$-clauses, $m_j$, for all $j\in [k]$.
		\If{$B\in\left[0,(1-\delta)(\sum_{j=2}^k \frac{2^j-j-1}{2^{j-2}}m_j) \right]$}
				\Statex {\bf Output: $v =  \sum_{j=1}^k (1-2^{-j})m_j +\frac{(1-\delta)^2B^2}{4\left(\sum_{j=2}^k (\frac{2^j-j-1}{2^{j-2}})m_j\right)}$}. 
		\Else
				\Statex {\bf Output: $v =  \sum_{j=1}^k \frac{j}{2^j} m_j +\frac{(1-\delta)B}{2}$}.
		\EndIf
	\end{algorithmic}
\end{algorithm}

\thmmksat*

\begin{proof}
We prove that Algorithm~\ref{alg:mSAT} computes a $\left(\frac{\sqrt{2}}{2}-\epsilon\right)$-approximation by showing that (i) $v\leq \val_\Psi$, and (ii) $v \geq \left( \frac{\sqrt{2}}{2} - \epsilon \right)\cdot\val_\Psi$, where $v$ is the output of the algorithm. Recall that by the guarantee of Theorem~\ref{thm:l1sketch}, with probability at least $3/4$:
\[
(1-\delta)\bias(\Psi) \leq B \leq (1+\delta)\bias(\Psi)
\, .
\]
\paragraph{(i) $\bm{v\leq \val_\Psi}$.}
If $B>(1-\delta)(\sum_{j=2}^k \frac{2^j-j-1}{2^{j-2}}m_j)$, then 
\[
v=\sum_{j=1}^k \frac{j}{2^j} m_j +\frac{(1-\delta)B}{2}
\leq \sum_{j=1}^k \frac{j}{2^j} m_j +\frac{(1-\delta^2)\bias(\Psi)}{2}
\leq \val_\Psi
\]
by the first bound in Lemma~\ref{lem:ksat bias lower bound}.

If $B\leq(1-\delta)(\sum_{j=2}^k \frac{2^j-j-1}{2^{j-2}}m_j)$, then $\bias(\Psi)\leq B/(1-\delta)\leq (\sum_{j=2}^k \frac{2^j-j-1}{2^{j-2}}m_j)$, and, thus, $\val_\Psi \geq
\sum_{j=1}^k (1-2^{-j})m_j +\frac{\bias(\Psi)^2}{4\left(\sum_{j=2}^k (\frac{2^j-j-1}{2^{j-2}})m_j\right)}$ by the second bound in Lemma~\ref{lem:ksat bias lower bound}. Then
\begin{align*}
v&= \sum_{j=1}^k (1-2^{-j})m_j +\frac{(1-\delta)^2 B^2}{4\left(\sum_{j=2}^k (\frac{2^j-j-1}{2^{j-2}})m_j\right)}\\
&\leq \sum_{j=1}^k (1-2^{-j})m_j+\frac{(1-\delta^2)\bias(\Psi)^2}{4\left(\sum_{j=2}^k (\frac{2^j-j-1}{2^{j-2}})m_j\right)}\leq \val_\Psi
\, .
\end{align*}

\paragraph{(ii) $\bm{v \geq \left(1 - \epsilon \right)\cdot \frac{\sqrt{2}}{2}  \val_\Psi}$.} 
We will in fact prove the stronger lower bound of $v \geq \left(1 - \epsilon \right)\cdot \frac{3}{4}  \val_\Psi$ when the bias is high, i.e., $B>(1-\delta)(\sum_{j=2}^k \frac{2^j-j-1}{2^{j-2}}m_j)$. We will prove the desired bound by considering the following two cases: (1) $m_1> \sum_{j=2}^k (4-(j+4)2^{1-j})m_j$, and (2) $m_1\le \sum_{j=2}^k (4-(j+4)2^{1-j})m_j$. If $m_1> \sum_{j=2}^k (4-(j+4)2^{1-j})m_j$, then
\[
B + m_1 \ge (1-\delta)\sum_{j=2}^k (8-(6j+12)2^{-j})m_j \ge (1-\delta) \sum_{j=2}^k (6-(2j+12)2^{-j})m_j
\, .
\]
Rearranging the terms, we get
\begin{align*}
v &= \sum_{j=1}^k \frac{j}{2^j} m_j +\frac{(1-\delta)B}{2} \\
&\ge (1-\delta)\left(\frac{B+m_1}{2}+\sum_{j=2}^k\frac{j}{2^j} m_j\right) \\
&\ge (1-\delta)\left(\frac{3}{4}\cdot\frac{B+m_1}{2}+\frac{B+m_1}{8}+\sum_{j=2}^k\frac{j}{2^j} m_j\right) \\
&\ge (1-\delta)^2 \cdot \frac{3}{4} \cdot \left(\frac{B}{2}+\sum_{j=1}^k (1+(j-2)2^{-j})m_j\right) \\
&\ge (1-\delta)(1-\delta^2) \cdot \frac{3}{4} \cdot \left(\frac{\bias(\Psi)}{2}+\sum_{j=1}^k (1+(j-2)2^{-j})m_j\right) \\
&\ge (1-\epsilon) \cdot \frac{3}{4} \cdot \val_{\Psi}\, ,
\end{align*}
by the bound in Lemma~\ref{lem:ubSAT}.
\\
\\
If $m_1\le \sum_{j=2}^k (4-(j+4)2^{1-j})m_j$, then
\begin{align*}
v &= \sum_{j=1}^k \frac{j}{2^j} m_j +\frac{(1-\delta)B}{2} \\
&\ge (1-\delta) \left(\sum_{j=1}^k  \frac{j}{2^j} m_j + \sum_{j=2}^k \frac{2^j-j-1}{2^{j-1}}m_j \right) \\
&= (1-\delta) \left(\frac{m_1}{2}+\sum_{j=2}^k \frac{3m_j}{4} + \sum_{j=2}^k (\frac{5}{4}-(j+2)2^{-j})m_j \right) \\
&= (1-\delta) \Bigg(\frac{m_1}{2}+\sum_{j=2}^k \frac{3m_j}{4} + \underbrace{\sum_{j=2}^k (1-(j/2+2)2^{-j})m_j}_{\ge m_1/4} +\underbrace{\sum_{j=2}^k(1/4-j\cdot2^{-(j+1)}) m_j}_{\ge 0} \Bigg)\\
&\ge (1-\delta) \cdot \frac{3}{4} \cdot \sum_{j=1}^k m_j \ge (1-\epsilon) \cdot \frac{3}{4} \cdot \val_{\Psi}
\, .
\end{align*}

We will now prove the lower bound of $v \geq \left(1 - \epsilon \right)\cdot \frac{\sqrt{2}}{2}  \val_\Psi$ when the bias is low, i.e., $B\leq(1-\delta)(\sum_{j=2}^k \frac{2^j-j-1}{2^{j-2}}m_j)$. We have the following two claims.

\begin{restatable}{claim}{maxksatmiddleB}
\label{claim:max k sat middle B}
If $\sum_{j=1}^k\frac{2-j}{2^{j-1}}m_j\leq B\leq(1-\delta)(\sum_{j=2}^k \frac{2^j-j-1}{2^{j-2}}m_j)$, then
\[
\sum_{j=1}^k(1-2^{-j})m_j+\frac{B^2}{4\sum_{j=2}^k\frac{2^j-j-1}{2^{j-2}}m_j}\geq\frac{\sqrt{2}}{2}\sum_{j=1}^km_j \, .
\]
\end{restatable}

\begin{restatable}{claim}{maxksatsmallB}
\label{claim:max k sat small B}
If $B\leq\sum_{j=1}^k\frac{2-j}{2^{j-1}}m_j$, then
\[
\sum_{j=1}^k(1-2^{-j})m_j+\frac{B^2}{4\sum_{j=2}^k\frac{2^j-j-1}{2^{j-2}}m_j}\geq\frac{\sqrt{2}}{2}\left(\frac{B}{2}+\sum_{j=1}^k\frac{2^j+j-2}{2^j}m_j\right) \, .
\]
\end{restatable}

Let us postpone the proof for the two claims to Section~\ref{sec:proof of max k sat middle B} and Section~\ref{sec:proof of max k sat small B} and complete the proof for Theorem~\ref{thm:mksat} assuming their correctness. If $\sum_{j=1}^k\frac{2-j}{2^{j-1}}m_j\leq B\leq(1-\delta)(\sum_{j=2}^k \frac{2^j-j-1}{2^{j-2}}m_j)$, then by Claim~\ref{claim:max k sat middle B} we have
\begin{align*}
v&=\sum_{j=1}^k(1-2^{-j})m_j+\frac{B^2}{4\sum_{j=2}^k\frac{2^j-j-1}{2^{j-2}}m_j}\geq\frac{\sqrt{2}}{2}\sum_{j=1}^km_j\\
&\geq\frac{\sqrt{2}}{2}\cdot\val_{\Psi} \, .
\end{align*}
If $B\leq\sum_{j=1}^k\frac{2-j}{2^{j-1}}m_j$, then by Claim~\ref{claim:max k sat small B}, we have
\begin{align*}
v=&\sum_{j=1}^k(1-2^{-j})m_j+\frac{B^2}{4\sum_{j=2}^k\frac{2^j-j-1}{2^{j-2}}m_j}\geq\frac{\sqrt{2}}{2}\left(\frac{B}{2}+\sum_{j=1}^k\frac{2^j+j-2}{2^j}m_j\right)\\
&\geq(1-\delta)\cdot\frac{\sqrt{2}}{2}\left(\frac{\bias(\Psi)}{2}+\sum_{j=1}^k\frac{2^j+j-2}{2^j}m_j\right)\\
(\because\text{Lemma~\ref{lem:ubSAT}})&\geq(1-\delta)\cdot\frac{\sqrt{2}}{2}\cdot\val_\Psi\geq(1-\epsilon)\cdot\frac{\sqrt{2}}{2}\cdot\val_\Psi \, .
\end{align*}
We conclude that $v\geq(1-\epsilon)\cdot\frac{\sqrt{2}}{2}\val_\Psi$. This completes the proof of Theorem~\ref{thm:mksat}.

\end{proof}

\subsection{Proof of Lemma~\ref{lem:ubSAT}}\label{sec:proofs_ksat ub}

\ubk*
\begin{proof}

Let $k$ be the length of the largest clause in $\Psi$. Since $\sum_{j=1}^k m_j$ is the number of clauses in~$\Psi$, the first bound $\val_{\Psi} \leq \sum_{j=1}^k m_j$ holds trivially. To show the second bound, we first negate all the variables of $\Psi$ with $\bias_i(\Psi)<0$. This transformation does not change $\bias(\Psi), \val_{\Psi}$, and $m_j$, for all $j$, and every assignment of the variables of the original instance can be uniquely mapped to a corresponding assignment for the new instance satisfying the same number of clauses.
Therefore, without loss of generality, for every $i\in[n]$,
\[
\sum_{j=1}^k \frac{1}{2^{j-1}} (\posvar_i^{(j)}(\Psi) - \negvar_i^{(j)}(\Psi))=\bias_i(\Psi)\geq0
\, .
\]
Consider an assignment $\sigma$ to the variables of $\Psi$. We need to show that $\val_\Psi(\sigma)\leq\frac{m_1+2m_2+\bias(\Psi)}{2} + \frac{9 m_{\ge 3}}{8}$. Let $T$ be the set of (indices of) variables of $\sigma$ assigned the value $1$. We denote by $S_j$ the number of $j$-clauses satisfied by $\sigma$. We will prove that the number of $k$-clauses satisfied by $\sigma$ is bounded by

\begin{align}
\label{eq:sk}
S_k \leq \min\left\{m_k, 2^{k-2}\bias(\Psi)+\sum_{j= 1}^k j\cdot 2^{k-j-1}\cdot m_j  - \sum_{j= 1}^{k-1} 2^{k-j} \cdot S_j \right\} \, .
\end{align}
First we show how~\eqref{eq:sk} finishes the proof of the lemma, and then prove~\eqref{eq:sk}.

Indeed, then the number of clauses satisfied by $\sigma$ is bounded from above by
\begin{align*}
\val_\Psi(\sigma) 
&\leq \sum_{j=1}^k S_j \\
&\leq \sum_{j=1}^{k-1} S_j + \min\left\{m_k, 2^{k-2}\bias(\Psi)+\sum_{j= 1}^k j\cdot 2^{k-j-1} \cdot m_j  - \sum_{j= 1}^{k-1} 2^{k-j} \cdot S_j \right\}\\
&\leq  \sum_{j= 1}^{k-1} S_j + \frac{2^{k-1}-1}{2^{k-1}}\cdot m_k +\frac{1}{2^{k-1}}\left(2^{k-2}\bias(\Psi)+\sum_{j= 1}^k j\cdot 2^{k-j-1}\cdot m_j  - \sum_{j= 1}^{k-1} 2^{k-j} \cdot S_j\right)\\
&= \frac{\bias(\Psi)}{2} + \sum_{j= 1}^k j\cdot 2^{-j} \cdot m_j + (1-2^{-k+1})m_k + \sum_{j= 1}^{k-1} (1-2^{-j+1})S_j\\
& \le \frac{\bias(\Psi)}{2} + \sum_{j= 1}^k (1+(j-2)2^{-j}) m_j
\, ,
\intertext{where the last inequality follows since $S_j$ is trivially upper bounded by $m_j$.}
\end{align*}

\end{proof}

\subsection{Proof of Lemma~\ref{lem:ksat bias lower bound}}\label{sec:proofs_ksat lb} 
\biaslbk*
\begin{proof}
Let $k$ be the length of the largest clause in $\Psi$. Without loss of generality, we assume that for every $i\in[n], \bias_i(\Psi)\geq 0$. (Again, we can negate all variables with $\bias_i(\Psi)<0$, and define a bijection between the assignments for the two formulas.)
Therefore, for every $i\in[n],$
\[
\sum_{j=1}^k \frac{1}{2^{j-1}} (\posvar_i^{(j)}(\Psi) - \negvar_i^{(j)}(\Psi))=\bias_i(\Psi)\geq0
\, .
\]
Consider the greedy assignment $\sigma$ which assigns $1$ to every variable. Under this assignment,
\[
\val_\Psi(\sigma) \ge \sum_{i\in[n]}\sum_{j= 1}^k \frac{\posvar_i^{(j)}(\Psi)}{j}
\, .
\]
Recall that
\begin{align*}
\bias(\Psi)
&= \sum_{i\in[n]} \sum_{j=1}^k \frac{1}{2^{j-1}} (\posvar_i^{(j)}(\Psi) - \negvar_i^{(j)}(\Psi))\, ,\\
m_j&= \sum_{i\in[n]}\frac{\posvar_i^{(j)}(\Psi)}{j}+\frac{\negvar_i^{(j)}(\Psi)}{j}\,
\, .
\end{align*}

It follows that
\[
\val_\Psi(\sigma) \ge \frac{\bias(\Psi)}{2} + \sum_{j=1}^k \frac{j}{2^j} m_j
\, .
\]
This proves the first inequality in Lemma~\ref{lem:ksat bias lower bound}.

To prove the second inequality in Lemma~\ref{lem:ksat bias lower bound}, we consider a distribution of assignments to the variables of $\Psi$, where every variable $x_i$ is assigned the value 1 independently with probability $(\frac{1}{2}+\gamma)$, for a parameter $\gamma\in[0, 1/2]$ to be assigned later. The expected number of satisfied $j$-clauses under this distribution is

\begin{align*}
S_j
&\ge \sum_{r=0}^j m_j^{(r)} \cdot \left(1- \left(\frac{1}{2}-\gamma\right)^{r}\left(\frac{1}{2}+\gamma\right)^{j-r} \right) \\
&\ge (1-2^{-j})\sum_{r=0}^j m_j^{(r)} + \left(\sum_{r=0}^j m_j^{(r)} \cdot \frac{2r-j}{2^{j-1}}\right)\gamma - \sum_{d=2}^{j} \sum_{r=0}^j m_j^{(r)} \cdot \frac{1}{2^{j-d}} \cdot \binom{j}{d} \gamma^d \\
&= (1-2^{-j})m_j + \left(\sum_{r=0}^j m_j^{(r)} \cdot \frac{2r-j}{2^{j-1}}\right)\gamma -  m_j \sum_{d=2}^{j} \frac{1}{2^{j-d}}\cdot \binom{j}{d} \gamma^d  \\
&= m_j + \left(\sum_{r=0}^j m_j^{(r)} \cdot \frac{2r-j}{2^{j-1}}\right)\gamma - m_j\left(\left(\frac{1}{2}+\gamma\right)^j - \frac{j}{2^{j-1}} \gamma \right) \\
&\ge m_j + \left(\sum_{r=0}^j m_j^{(r)} \cdot \frac{2r-j}{2^{j-1}}\right)\gamma - m_j \left(2^{-j} + \frac{2^j-j-1}{2^{j-2}}\gamma^2\right)
\, ,
\end{align*}
where the last inequality follows by applying the following inequality which holds for all $0\le x\le 1$
\[
(1+x)^n \le 1 + nx + (2^n-n-1)x^2
\, .
\]
Let $m_j^{(r)}$ denote the number of $j$-clauses with $r$ positive literals. Note that a $j$-clause with $r$ positive literals contributes $\frac{2r-j}{2^{j-1}}$ to the total bias. Therefore,
\[
\bias(\Psi)
= \sum_{j= 1}^k \sum_{r=0}^j \frac{2r-j}{2^{j-1}} \cdot m_j^{(r)}
\, .
\]

Let us now compute the expected number of clauses satisfied by an assignment $\sigma$ from the distribution defined above.
\begin{align*}
\Exp_{\sigma}\left[\val_\Psi(\sigma)\right] 
= \sum_{j= 1}^k S_j
&\geq \sum_{j= 1}^k (1-2^{-j})m_j +  \gamma\cdot\left(\sum_{j= 1}^k\sum_{r=0}^j m_j^{(r)} \cdot \frac{2r-j}{2^{j-1}}\right) -\gamma^2 \sum_{j=2}^k \frac{2^j-j-1}{2^{j-2}}m_j \\
&=\sum_{j= 1}^k (1-2^{-j})m_j +  \gamma\cdot\bias(\Psi) -\gamma^2 \sum_{j=2}^k \frac{2^j-j-1}{2^{j-2}}m_j
\, .
\end{align*}

For the case where $\bias(\Psi)\leq \sum_{j=2}^k \frac{2^j-j-1}{2^{j-2}} m_j$, we set
$\gamma =\frac{\bias(\Psi)}{2\left(\sum_{j=2}^k \frac{2^j-j-1}{2^{j-2}} m_j\right)}\in[0,1/2]$, and derive the second bound:
\begin{align*}
\val_\Psi\geq\Exp_{\sigma}\left[\val_\Psi(\sigma)\right] &\geq \sum_{j= 1}^k (1-2^{-j})m_j + \frac{\bias(\Psi)^2}{2\left(\sum_{j=2}^k \frac{2^j-j-1}{2^{j-2}} m_j\right)} - \frac{\bias(\Psi)^2}{4\left(\sum_{j=2}^k \frac{2^j-j-1}{2^{j-2}} m_j\right)^2} \left(\sum_{j=2}^k \frac{2^j-j-1}{2^{j-2}} m_j\right) \\
&\ge  \sum_{j=1}^k (1-2^{-j})m_j +\frac{\bias(\Psi)^2}{4\left(\sum_{j=2}^k (\frac{2^j-j-1}{2^{j-2}})m_j\right)} \, .
\end{align*}

\end{proof}

\subsection{Proof of Claim~\ref{claim:max k sat middle B}}\label{sec:proof of max k sat middle B}
We will use the following claim.
\begin{claim}
\label{claimnew}
For every $x\geq 0, y>0, y\geq a\geq 0$:
\[
\frac{2x+3y+x^2/y-2a}{4(x+y)-3a}\geq\frac{\sqrt{2}}{2}
\, .
\]
\end{claim}
\begin{proof}
Let $L=2x+3y+x^2/y$ and $M=4(x+y)>3a$. From Claim~\ref{claim}, $L/M\geq\frac{\sqrt{2}}{2}>\frac{2}{3}$.
First we observe that 
\[
\frac{L-2a}{M-3a}-\frac{L}{M}=\frac{a(3L-2M)}{M(M-3a)}\geq 0
\]
for every $a\geq0$ and $M>3a$.
Now
\[
\frac{2x+3y+x^2/y-2a}{4(x+y)-3a}
=\frac{L-2a}{M-3a} \geq \frac{L}{M} \geq \frac{\sqrt{2}}{2} \,,
\]
where the last inequality uses Claim~\ref{claim} again.
\end{proof}

We are now ready to prove Claim~\ref{claim:max k sat middle B}.
\maxksatmiddleB*
\begin{proof}
Let
\[
V=\frac{\sum_{j=1}^k\frac{2^j-1}{2^{j-2}}m_j+\frac{B^2}{\sum_{j=2}^k\frac{2^j-j-1}{2^{j-2}}m_j}}{4\sum_{j=1}^km_j} \, .
\]
Now it remains to show that $V\geq\sqrt{2}/2$. First, observe that
\[
V\geq\frac{\sum_{j=1}^k(4-2^{2-j})m_j}{4\sum_{j=1}^km_j}\geq\frac{1}{2} \, .
\]
Also, if $x/y\geq1/2$ for non-negative $x,y$, then we have $x/y\geq(x+a)/(y+2a)$ for all $a\geq0$. As $B\geq\sum_{j=1}^k\frac{2-j}{2^{j-1}}m_j$, we can lower bound $V$ as follows.
\begin{align*}
V&=\frac{\sum_{j=1}^k\frac{2^j-1}{2^{j-2}}m_j+\frac{B^2}{\sum_{j=2}^k\frac{2^j-j-1}{2^{j-2}}m_j}}{4\sum_{j=1}^km_j}\\
&\geq\frac{\sum_{j=1}^k\frac{2^j-1}{2^{j-2}}m_j+\frac{B^2}{\sum_{j=2}^k\frac{2^j-j-1}{2^{j-2}}m_j}+2(B-\sum_{j=1}^k\frac{2-j}{2^{j-1}}m_j)}{4\sum_{j=1}^km_j+4(B-\sum_{j=1}^k\frac{2-j}{2^{j-1}}m_j)}\\
&=\frac{2B+\sum_{j=1}^k\frac{2^j+j-3}{2^{j-2}}m_j+\frac{B^2}{\sum_{j=2}^k\frac{2^j-j-1}{2^{j-2}}m_j}}{4B+\sum_{j=1}^k\frac{2^j+2j-4}{2^{j-2}}m_j} \, .
\intertext{Now, let $x=B$,  $y=\sum_{j=2}^k\frac{2^j-j-1}{2^{j-2}}m_j$, and $a=\sum_{j=1}^k\frac{2^j-2j}{2^{j-2}}m_j$, and rewrite the above quantity as follows}
&=\frac{2x+3y+x^2/y-2a}{4(x+y)-3a} \\
&\geq \frac{\sqrt{2}}{2}\, ,
\end{align*}
where the last inequality follows from Claim~\ref{claimnew} and $y\geq a$.
\end{proof}

\subsection{Proof of Claim~\ref{claim:max k sat small B}}\label{sec:proof of max k sat small B}
\maxksatsmallB*
\begin{proof}
Let
\[
V=\frac{\sum_{j=1}^k\frac{2^j-1}{2^{j-2}}m_j+\frac{B^2}{\sum_{j=2}^k\frac{2^j-j-1}{2^{j-2}}m_j}}{2B+\sum_{j=1}^k\frac{2^j+j-2}{2^{j-2}}m_j} \, .
\]
Now it remains to show that $V\geq\sqrt{2}/2$. Let $R=\sum_{j=1}^k\frac{2-j}{2^{j-2}}m_j-2B\geq0$. Since the denominator of $V$ is not less than both $R$ and the numerator of $V$, we have that 
\begin{align*}
V&\geq
\frac{\sum_{j=1}^k\frac{2^j-1}{2^{j-2}}m_j+\frac{B^2}{\sum_{j=2}^k\frac{2^j-j-1}{2^{j-2}}m_j}-R}{2B+\sum_{j=1}^k\frac{2^j+j-2}{2^{j-2}}m_j-R}\\
&=\frac{\sum_{j=1}^k\frac{2^j-1}{2^{j-2}}m_j+\frac{B^2}{\sum_{j=2}^k\frac{2^j-j-1}{2^{j-2}}m_j}-(\sum_{j=1}^k\frac{2-j}{2^{j-2}}m_j-2B)}{2B+\sum_{j=1}^k\frac{2^j+j-2}{2^{j-2}}m_j-(\sum_{j=1}^k\frac{2-j}{2^{j-2}}m_j-2B)}\\
&=\frac{2B+\sum_{j=1}^k\frac{2^j+j-3}{2^{j-2}}m_j+\frac{B^2}{\sum_{j=2}^k\frac{2^j-j-1}{2^{j-2}}m_j}}{4B+\sum_{j=1}^k\frac{2^j+2j-4}{2^{j-2}}m_j} \, .
\intertext{Now, let $x=B$, $y=\sum_{j=1}^k\frac{2^j-j-1}{2^{j-2}}m_j$, and $a=\sum_{j=1}^k\frac{2^j-2j}{2^{j-2}}m_j$, and rewrite the above quantity as follows.}
&=\frac{2x+3y+x^2/y-2a}{4(x+y)-3a} \\
&\geq \frac{\sqrt{2}}{2}\, ,
\end{align*}
where the last inequality follows from Claim~\ref{claimnew} and $y\geq a$.
\end{proof}

%% file: future.tex
\section*{Open Questions}
Our work gives optimal approximation ratios for all Boolean maximum constraint satisfaction problems with constraints of length at most two. It would be interesting to understand the complexity of constraint languages with \emph{arity greater than two}, and \emph{larger alphabet} sizes.

In terms of lower bounds, we show that better than $\frac{4}{9}$- and $\frac{\sqrt{2}}{2}$-approximations for \textsf{Max-2-AND} and \textsf{Max-2-OR} require space $\Omega(\sqrt{n})$. Can we improve these space lower bounds to $\Omega(n)$, matching the space requirements of standard algorithms that give $(1-\epsilon)$-approximation?

%% file: acknowledgement.tex
\subsection*{Acknowledgement}
We thank Madhu Sudan for many helpful discussions, and for pointing out a mistake in an old proof. We also thank Mitali Bafna for spotting an error in our previous algorithm for $\mksat$. Finally, we thank the anonymous referees for their several useful suggestions which helped improve the presentation of the paper.